\documentclass[a4paper,english,aps,reprint]{revtex4-1}
\usepackage{lmodern}
\usepackage{lmodern}
\usepackage[T1]{fontenc}
\usepackage[utf8]{inputenc}
\usepackage{babel}
\usepackage{verbatim}
\usepackage{amsmath}
\usepackage{amssymb}
\usepackage{stackrel}
\usepackage{graphicx}
\usepackage{esint}
\usepackage[unicode=true,pdfusetitle,
 bookmarks=true,bookmarksnumbered=false,bookmarksopen=true,bookmarksopenlevel=1,
 breaklinks=false,pdfborder={0 0 0},backref=false,colorlinks=false]
 {hyperref}
\usepackage{xcolor}

\makeatletter

\pdfpageheight\paperheight
\pdfpagewidth\paperwidth

\@ifundefined{textcolor}{}
{%
 \definecolor{BLACK}{gray}{0}
 \definecolor{WHITE}{gray}{1}
 \definecolor{RED}{rgb}{1,0,0}
 \definecolor{GREEN}{rgb}{0,1,0}
 \definecolor{BLUE}{rgb}{0,0,1}
 \definecolor{CYAN}{cmyk}{1,0,0,0}
 \definecolor{MAGENTA}{cmyk}{0,1,0,0}
 \definecolor{YELLOW}{cmyk}{0,0,1,0}
}

\usepackage{babel}
\usepackage{babel}
\usepackage{babel}
\usepackage{babel}
\usepackage{babel}
\usepackage[normalem]{ulem}
\usepackage{babel}
\usepackage{babel}
\usepackage{amsfonts}
\providecommand{\U}[1]{\protect\rule{.1in}{.1in}}
\pdfpageheight\paperheight
\pdfpagewidth\paperwidth

\@ifundefined{textcolor}{}{
\definecolor{BLACK}{gray}{0}
\definecolor{WHITE}{gray}{1}
\definecolor{RED}{rgb}{1,0,0}
\definecolor{GREEN}{rgb}{0,1,0}
\definecolor{BLUE}{rgb}{0,0,1}
\definecolor{CYAN}{cmyk}{1,0,0,0}
\definecolor{MAGENTA}{cmyk}{0,1,0,0}
\definecolor{YELLOW}{cmyk}{0,0,1,0}
}
\pdfpageheight\paperheight
\pdfpagewidth\paperwidth

\@ifundefined{textcolor}{}{
\definecolor{BLACK}{gray}{0}
\definecolor{WHITE}{gray}{1}
\definecolor{RED}{rgb}{1,0,0}
\definecolor{GREEN}{rgb}{0,1,0}
\definecolor{BLUE}{rgb}{0,0,1}
\definecolor{CYAN}{cmyk}{1,0,0,0}
\definecolor{MAGENTA}{cmyk}{0,1,0,0}
\definecolor{YELLOW}{cmyk}{0,0,1,0}
}

\makeatother

\begin{document}

\title{Density-based one-dimensional model potentials for strong-field\\
 simulations in $\text{He}$, $\text{H}_{2}^{+}$ and $\text{H}_{2}$}

\author{Szilárd Majorosi\textsuperscript{1}, Mihály G. Benedict\textsuperscript{1},
Ferenc Bogár\textsuperscript{2}, Gábor Paragi\textsuperscript{2,3}
and Attila Czirják\textsuperscript{1,4,}}

\email{czirjak@physx.u-szeged.hu}

\affiliation{\textsuperscript{1}Department of Theoretical Physics, University
of Szeged, Tisza L. krt. 84-86, H-6720 Szeged, Hungary\linebreak{}
 \textsuperscript{2}MTA-SZTE Biomimetic Systems Research Group, University
of Szeged, Dóm tér 8, H-6720 Szeged, Hungary \linebreak{}
 \textsuperscript{3}Institute of Physics, University of Pécs, Ifjúság
útja 6, H-7624 Pécs, Hungary \linebreak{}
 \textsuperscript{4}ELI-ALPS, ELI-HU Non-Profit Ltd., Dugonics tér
13, H-6720 Szeged, Hungary}

\begin{abstract}

We present results on the accurate one-dimensional (1D) modeling of simple atomic
and molecular systems excited by strong laser fields. We use
atomic model potentials that we derive from the corrections proposed earlier
using the reduced ground state density of a three-dimensional (3D)
single-active electron atom.
The correction involves a change of the asymptotics
of the 1D Coulomb model potentials while maintaining the correct
ground state energy.
We present three different applications of this
method: we construct correct 1D models of the hydrogen
molecular ion, the helium atom and the hydrogen molecule using improved
parameters of existing soft-core Coulomb potential forms. We test
these 1D models by comparing the corresponding numerical simulation results with
their 3D counterparts in typical strong-field physics scenarios 
with near- and mid-infrared laser pulses, having peak intensities in the $10^{14}-10^{15}\,\mathrm{W/cm}^2$ range,
and we find
an impressively increased accuracy in the dynamics of
the most important atomic quantities on the time scale of the excitation.
We also present the high-order harmonic spectra of the He atom, computed using
our 1D atomic model potentials.
They show a very good match with the structure and phase obtained
from the 3D simulations in an experimentally important range of excitation
amplitudes.

\end{abstract}
\maketitle

\section{Introduction\label{sec:introduction}}

The interpretation of typical experiments in attosecond and strong-field physics, including the pioneering results in Refs. \cite{hentschel2001attosecond,kienberger2002attosecond,drescher2002attosecond,baltuska2003attosecond,Uiberacker_Nature_2007,hommelhoff2009extremelocalization,Schultze_Science_2010,Haessler_NatPhys_2010,pfeiffer2012attoclock,shafir2012tunneling,ranitovic2014attosecondcontrol,ciappina2017attosecondnano,AzouryKrugerDudovich2017SelfprobingXUVphotoionization},
often relies on the quantum description of the involved atomic system
driven by a strong laser pulse \cite{Keldysh_JETP_1965,varro1993multiphotonequation,lewenstein1994hhgtheory,protopapas1997tdseionization,ivanov2005strongfield,gordon2005tdsecoulomb,Krausz_RevModPhys_2009_Attosecond_physics,frolov2012attosecondanalytic,peng2015attosecondtracing}.
Despite recent progress in analytic and numerical solution techniques
\cite{clarke2018rmatrixultrafast,jiang2017splitlanczos,kidd2017exptddft,dijk2017tdsenumeric2d,LacknerBurgdoefer2017hhg_2p_rdm,
Kasza2018Sturmian,Efimov2018Restricted_ionization,Patchkovskii_CPC_2016},
the exact solution of the corresponding true 3D Schrödinger equation
is beyond reach in this non-perturbative range (except for the simplest
cases), which justifies the importance of good approximations.

If the strong driving laser pulse is linearly polarized then the most
important features of the resulting quantum dynamics can usually be
captured by a one-dimensional (1D) approximation \cite{eberly1988softcoulombspecra,eberly1991softcoulombatom,bauer1997tdse1dhemodel,chirilua2010HHGemission,silaev2010tdsecoulombs,sveshnikov2012schrodingercoulomb,graefe2012quantumphasespace,czirjak2000ionizationwigner,benedict2012entanglementtwo,czirjak2013rescatterentanglement,geltman2011boundstatesdelta,PhysRevB.88.075438,baumann2015wignerionization,teeny2016ionizationtime,Nishi_PhysRevA_100_013421_2019}.
Such 1D models have increasing importance for longer laser wavelengths where typical 3D simulations become inefficient 
\cite{Thumm_2014_PhysRevA.89.063423_3d-tdse-midIR,Madsen_2016_PhysRevA.93.033420_3d-tdse-midIR-speedup}.
These typically use various 1D model potentials to account for the
motion of the atomic system along the direction of the laser polarization.
However, the particular model potential can strongly influence some
of the 1D results and their quantitative comparison with the true
three-dimensional (3D) results is usually non-trivial \cite{bandrauk2009tdsehydrogen,graefe2012quantumphasespace,majorosi2017entanglement,IvanovIA2019Entropy_view_sfi}.

We addressed this problem for a single active electron atom in Ref.
\cite{majorosi2018densitymodel}: we introduced the density-based
1D model potential and, based on its features, we also found improved
parameters for other well-known 1D model potentials. The promising
strong-field simulation results inspired us to extend our modeling
approach to simple atomic systems like the hydrogen molecular ion,
the helium atom and the hydrogen molecule, which is the subject of
the present paper. Our key idea is to require the Coulomb asymptotes
in the 1D model potentials to be equal to those obtained from the
corresponding reduced 3D ground state single-electron density along
the direction of the laser polarization. Then we present the results
of careful numerical simulations of strong-field ionization scenarios
using these 1D model potentials,
considering nearly single-cycle laser pulses with carrier wavelengths of 725 nm and 3045 nm. 
Comparing them with the corresponding
3D simulation results, we make a conclusion about the recommended
use of these 1D model systems. We use atomic units in this paper.

\section{3D reference systems\label{sec:3D-model-systems}}

In this section, we specify in more detail the strong-field modeling
of the selected three-dimensional systems: the helium atom, the hydrogen
molecular ion and the hydrogen molecule 
driven by a linearly polarized laser pulse. We also outline the
underlying numerical simulations, the results of which we use later
as reference when we compare the corresponding one-dimensional results.

Although a suitable laser pulse may create also vibrations
and rotations in a diatomic molecule, 
we assume the nuclear motion to be frozen throughout this paper, 
and we set the molecular axis parallel to the polarization of the laser pulse.

For the H$_{2}^{+}$, we solve the three-dimensional Schrödinger equation,
both to compute the ground state, and to obtain the time-evolution
when driven by the laser pulse.  
For the two-electron systems, He and
H$_{2}$, we chose the time-dependent Hartree-Fock (TDHF) approach as the
reference model, using a single atomic orbital in real three-dimensional
space. 
According to Ref. \cite{Sato_PhysRevA_94_023405_2016}, this method provides a good approximation of the more elaborate and numerically demanding 
two-electron results for He, driven by a laser pulse with parameters very close to those in the present work,
since the effect of electron correlation is relatively small for He.
We assume that the ground states of He and H$_{2}$
are spin singlets 
and that the laser pulse does not interact with the spin degrees of freedom,
thus the orbital part of the two-electron wave function remains
symmetric during the time evolution.

The governing equation of the electrons' motion can be cast for all
of the above cases into the following 
form, using cylindrical coordinates $\rho=\sqrt{x^{2}+y^{2}}$ and
$z$: 
\begin{equation}
i\frac{\partial\Psi^{\mathrm{3D}}}{\partial t}=\left[T_{z}+T_{\rho}+V+z\mathcal{E}_{z}(t)+(N-1)V_{\mathrm{H}}\right]\Psi^{\mathrm{3D}},\label{eq:3d_tdse}
\end{equation}
 where 
 the kinetic energy operator is split as 
\begin{equation}
T_{z}=-\frac{1}{2}\frac{\partial^{2}}{\partial z^{2}},\,\,\,\,\,\,\,\,\,\, T_{\rho}=-\frac{1}{2}\left[\frac{\partial^{2}}{\partial\rho^{2}}+\frac{1}{\rho}\frac{\partial}{\partial\rho}\right].\label{eq:3d_kinetic_z_rho}
\end{equation}
The one-electron potential 
\begin{equation}
V(z,\rho)=-\frac{1}{\sqrt{\rho^{2}+(z-\frac{d}{2})^{2}}}-\frac{1}{\sqrt{\rho^{2}+(z+\frac{d}{2})^{2}}}\label{eq:3d_pot_single}
\end{equation}
contains the Coulomb interaction with the nuclei. The parameter $d$
is the internuclear distance for $\mathrm{H}_{2}^{+}$ and $\mathrm{H}_{2}$,
while for $d=0$ we get the helium atom (with its nucleus in the origin).
The $z\mathcal{E}_{z}(t)$ term of \eqref{eq:3d_tdse} corresponds to the interaction
with the laser field, polarized along the $z$-axis, using
dipole approximation and length gauge
 \cite{BOOK_QUANTUM_GRIFFITS_2005,BOOK_ATOMS_MOLECULES_BRANSDEN_2003,PhysRevB.96.035112}.
The $\mathcal{E}_{z}(t)$ denotes the electric field of the laser pulse evaluated
in the origin, and we assume that it is present only after $t>0$,
i.e. $\mathcal{E}_{z}(t\leq0)=0$. 

In \eqref{eq:3d_tdse}, we distinguish the one- and the two-electron
cases by the parameter$\ N$. For $N=2$ the electron-electron interaction
is described in \eqref{eq:3d_tdse} by the time-dependent Hartree-potential
 which is  given by 
\begin{equation}
V_{\mathrm{H}}(\mathbf{r},t;\Psi^{\mathrm{3D}})=\int\frac{|\Psi^{\mathrm{3D}}(\mathbf{r^{\prime}},t)|^{2}}{|\mathbf{r}-\mathbf{r}^{\prime}|}\mathrm{d}^{3}\mathbf{r}^{\prime}.\label{eq:3d_pot_hartree}
\end{equation}
The presence of this potential makes equation \eqref{eq:3d_tdse}
nonlinear in $\Psi^{\mathrm{3D}}$. In actual computations, we obtain this potential
by solving the corresponding discretized Poisson equation $\nabla^{2}V_{\mathrm{H}}=-4\pi|\Psi^{\mathrm{3D}}(\mathbf{r},t)|^{2}$
\ in cylindrical coordinates, to avoid the high dimensional integration.

This time-dependent Hamiltonian in \eqref{eq:3d_tdse} has axial symmetry
around the polarization axis of the electric field of the laser pulse
which makes the use of cylindrical coordinates practical, and provides
efficient calculation of the reduced dynamics along the $z$-axis.

For actual simulations, we use the efficient numerical method described
in \cite{majorosi2016tdsesolve}, which incorporates the singularity
of the Hamiltonian directly, using the required discretized Neumann
and Robin boundary conditions. We compute the ground states via imaginary
time propagation with high-order split-operator approximations \cite{chin2009imagsplit},
then we compute the time evolution up to a specified time $T_{\max}$.

To characterize the effects of the external field, we shall \ use
the ground state population loss in a single-electron wavefunction
or in an electron-orbital, defined as 
\begin{equation}
g(t)=1-\left\vert \left\langle \Psi^{\mathrm{3D}}(t=0)|\Psi^{\mathrm{3D}}(t)\right\rangle \right\vert ^{2},\label{eq:proj3_ground_loss}
\end{equation}
which is to be compared with the corresponding function in the 1D\ models
\ we consider below.

In our considerations, the electron density given by 
\begin{equation}
\varrho^{\mathrm{3D}}(z,\rho,t)=N\left\vert \Psi^{\mathrm{3D}}(z,\rho,t)\right\vert ^{2},\label{eq:3d_density}
\end{equation}
plays an important role as we shall construct our one dimensional model potentials by an appropriate reduction of this quantity.

\section{1D model systems\label{sec:1D-model-systems}}

In order to model the above described 3D strong-field process in 1D,
it is customary to use the following form of the time-dependent 1D
Hamiltonian 
\begin{equation}
H(t)=H_{0}+z\mathcal{E}_{z}(t)=T_{z}+V_{0}(z)+z\mathcal{E}_{z}(t).\label{eq:1d_hamiltonian}
\end{equation}
where the effects of a strong few-cycle laser pulse are to be modeled by
the very same electric field $\mathcal{E}_{z}(t)$ as in 3D.

We want to verify the physical correctness of the above models by
numerically solving the time-dependent Schrödinger equation 
\begin{equation}
i\frac{\partial}{\partial t}\Psi\left(z,t\right) = H(t)
\Psi\left(z,t\right)\label{eq:1d_tdse}
\end{equation}
and compare the time-dependent physical response of this system
 with that of the original 3D TDSE.

\subsection{Overview of 1D density-based atomic model potentials\label{subsec:model_1d_atom}}

The main question in (\ref{eq:1d_hamiltonian}) is the form of $V_{0}(z)$
\cite{eberly1988softcoulombspecra,eberly1991softcoulombatom,bauer1997tdse1dhemodel,chirilua2010HHGemission,silaev2010tdsecoulombs,sveshnikov2012schrodingercoulomb,graefe2012quantumphasespace,czirjak2000ionizationwigner,czirjak2013rescatterentanglement,geltman2011boundstatesdelta,baumann2015wignerionization,teeny2016ionizationtime}.
One of the possibilities proposed in our previous work \cite{majorosi2018densitymodel}
for \emph{one electron} atomic systems was that we have introduced
the \emph{reduced} one dimensional ground state density depending
only on the $z$ coordinate as 
\begin{equation}
\varrho_{z}(z)=2\pi\int\varrho^{\mathrm{3D}}(z,\rho)\,\rho\,\mathrm{d}\rho,\label{eq:3d_density_z}
\end{equation}
where $\varrho^{\mathrm{3D}}(z,\rho)$ is the 3D ground state density.
This made it possible to calculate the density based model potential
\begin{equation}
V_{0}(z)=E_{0}+\frac{1}{\sqrt{\varrho_{z}(z)}}T_{z}\sqrt{\varrho_{z}(z)},\label{eq:1d_density_pot}
\end{equation}
where $\sqrt{\varrho_{z}(z)}$ stands as the reduced \emph{ground
state} wavefunction, while $E_{0}$ is the corresponding exact 3D
ground state energy. This construction ensures that the reduced problem
yields exactly the same properties for the ground state as does the
original 3D atomic calculation. It is also an important feature that
this form preserves the ground state energy in our case, since the
original 3D problem has a long-range Coulomb asymptotics.

In addition to these important physical properties, in the cases of
single-electron atoms an analytic expression could be calculated for
$V_{0}(z)$ \cite{majorosi2018densitymodel}, which is of the form
of a short-range correction plus a 1D regularized Coulomb potential.
This long range Coulomb-part had an asymptotic form of $-\frac{1}{2}Z/z$,
where $Z$ is the nuclear charge in 3D. This inspired us to develop
\cite{majorosi2018densitymodel} an alternative, improved soft-core
Coulomb potential form, which is smooth and easy-to use: 
\begin{equation}
V_{0,\mathrm{Sc}}(z)=-\frac{Z^{\ast}}{\sqrt{z^{2}+ \left[ \alpha \left( Z^{\ast}\right) \right]^{2}}}\label{eq:1d_pot_soft}
\end{equation}
Here the parameter $Z^{\ast}$ in the numerator has been determined
by requiring to obtain the same asymptotic behavior as that of \eqref{eq:1d_density_pot},
while $\alpha$ is a fitting parameter depending on $Z^{\ast}$,
set by demanding the 1D atomic system to have the same ground state
energy as the corresponding 3D system. In case of single-electron
atoms, the parameters turned out to be \cite{majorosi2018densitymodel}
\begin{equation}
Z^{\ast}=\frac{Z}{2},\,\,\,\,\,\alpha^{2}=\frac{1}{16(Z^{\ast})^{2}}=\frac{1}{4Z^{2}}\,\,\,\,\,\text{with \thinspace\thinspace\thinspace\thinspace\thinspace}E_{0}=-\frac{Z^{2}}{2}.\label{eq:1d_params_soft}
\end{equation}
By using these potentials in  the solutions of the corresponding one
dimensional time-dependent Schrödinger equation \eqref{eq:1d_tdse}
 the results reproduced the 3D system's strong-field response quantitatively
correctly for several physical quantities. A detailed comparison can
be also found in \cite{majorosi2018densitymodel}. In those tests
the soft-core Coulomb form of the atomic model potential gave the
best physical response, despite yielding a slightly less accurate
ground state density.

It has been also found that even the power spectrum $p^{\mathrm{3D}}(f)$
of coherent high-order-harmonic generation obtained from 3D simulations
could be recovered from the corresponding 1D spectrum $p(f)$ by scaling
the latter as $p(f)/s(f),$ where $s(f)$ is the simple scaling function
\begin{equation}
s(f)=\min\left(1+0.03\left(100f-1\right)^{2},1+\left\vert 100f-1\right\vert \right),\label{eq:1d_power_scale}
\end{equation}
which turned out to be essentially independent from the strength and
the form of the exciting pulse $\mathcal{E}_{z}(t).$

The one-dimensional models we introduced in \cite{majorosi2018densitymodel}
proved to be more than simple toy models, by their capability of providing
quantitatively comparable results to the 3D ones. In this article
we test the physical relevance of these one-dimensional models by
extending them to simple composite atomic and molecular systems like
the one-electron diatomic molecule H$_{2}^{+}$, as well as to two
two-electron systems namely the He atom and the $\mathrm{H}_{2}$
molecule.

\subsection{1D hydrogen molecular ion model\label{subsec:model_1d_h2+}}

In our 1D model of the H$_{2}^{+}$ we assume that the molecular axis
is set along the $z$-direction and the internuclear distance $d$
is fixed (i.e. we do not consider nuclear motion). We are interested
in the strong-field dynamics of the electron according to the 1D Schrödinger
equation with the Hamiltonian (\ref{eq:1d_hamiltonian}). For the
1D model potential $V_{0}(z)$ in (\ref{eq:1d_hamiltonian}), we suggest
and test two candidates, $V_{0}^{(\mathrm{M})}$ and $V_{0,\mathrm{Sc}}^{(\mathrm{M})}$
as follows.

Based on our earlier results, summarized in the previous section,
we define the density-based hydrogen molecular ion potential on an
equidistant grid $z_{i}$ as 
\begin{equation}
V_{0}^{(\mathrm{M})}(z_{i};d)=E_{0}+\frac{1}{\sqrt{\varrho_{z}(z_{i};d)}}\widetilde{T}_{z}\sqrt{\varrho_{z}(z_{i};d)},\label{eq:1d_density_pot_mol}
\end{equation}
where $\varrho_{z}(z_{i};d)$ is the reduced density of the $1\sigma_{g}$
ground state of a 3D hydrogen molecular ion (with a fixed internuclear
distance $d$). This potential is calculated numerically with the
finite-difference version of the kinetic energy operator, denoted by $\widetilde{T}_{z}$
in (\ref{eq:1d_density_pot_mol}), and 
using the numerically exact ground
state energy which equals that of the 3D reference system \cite{majorosi2018densitymodel}.

\begin{figure}[h]
\includegraphics[width=1\columnwidth]{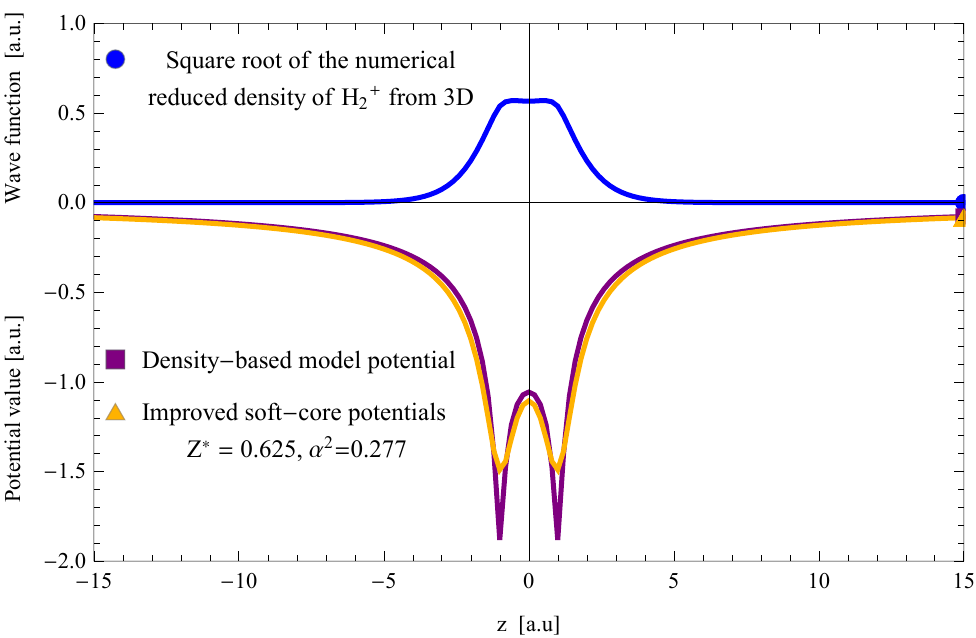}\protect\caption{Upper panel: the plot of the square root of reduced density of the
hydrogen molecular ion with $d=2.0$ (in blue). Lower panel: the plot
of the density-based model potential \eqref{eq:1d_density_pot_mol}
(in purple) and the improved soft-core Coulomb model \eqref{eq:1d_soft_pot_mol}
with the indicated parameter values 
(in gold). 
The ground state energy is $E_{0}=-1.1026$ for both of the potentials, 
using $\Delta z=0.2$ in \eqref{eq:1d_density_pot_mol}.}

\label{fig:h2+_config}
\end{figure}

The other 1D model potential we propose to use for the H$_{2}^{+}$
is the soft-core molecular model potential 
\begin{equation}
V_{0,\mathrm{Sc}}^{(\mathrm{M})}(z;d)=V_{0,\mathrm{Sc}}\left(z-\frac{d}{2}\right)+V_{0,\mathrm{Sc}}\left(z+\frac{d}{2}\right)\label{eq:1d_soft_pot_mol}
\end{equation}
where we assume an implicit dependence of the parameters $Z^{\ast}$
and $\alpha^{2}$ in $V_{0,\mathrm{Sc}}$ on the parameter $d$. 
The value of $Z^{\ast}$ determining the Coulomb asymptotics of \eqref{eq:1d_soft_pot_mol}
is calculated from the potential given by \eqref{eq:1d_density_pot_mol}.
Then, $\alpha^{2}$ can be determined by setting the correct single-electron
energy from the reference 3D hydrogen molecular ion calculation. We
plot the shape of the corresponding 1D potentials with $d=2$, along
with $\sqrt{\varrho_{z}(z_{i};d=2.0})$ in Fig. \ref{fig:h2+_config}
as well as the values of the parameters $Z^{\ast}$ and $\alpha^{2}$
obtained in this way.

In these computations, we have typically employed a grid spacing
of $\Delta z=0.2$, which yields about 3-4 digit accuracy in the ground
state density of the reference 3D method. For an intermolecular distance
$d=2.0$ -- that is near the equilibrium distance of the hydrogen
molecular ion -- the ground state energy is $E_{0}=-1.1026$ based
on both our 3D reference calculation and the 1D model using \eqref{eq:1d_density_pot_mol}.
Schaad and Hicks \cite{schaad1970h2+bond} give a very accurate result
for the equilibrium distance as $d=1.9972$, and an electron energy
of $E_{0}=-1.10334$. (The binding energy they gave was $-0.602634619$
a.u. which also incorporated the $1/d$ \ Coulomb repulsion energy
of the protons.) 

\subsection{1D helium atom model\label{subsec:model_2d_helium}}

We turn now to model a two-electron system, the helium atom in one
spatial dimension denoting the electron coordinates by $z_{1}$ and
$z_{2}$. The key element of our model is that we replace the true
Coulomb potentials with 1D improved soft-core potentials $V_{0,\mathrm{Sc}}$
of the form \eqref{eq:1d_pot_soft}, both for the
electron-nucleus and for the electron-electron interaction. 
By setting $Z^{\ast}=1$ in $V_{0,\mathrm{Sc}}$, we have
the following Hamiltonian:
\begin{equation}
H_{\mathrm{He}}^{\mathrm{1D}}=\sum_{k=1}^{2}\left[T_{z_{k}}+V_{0,\mathrm{Sc}}(z_{k})+z_{k}\mathcal{E}_{z}(t)\right]-\frac{1}{2}V_{0,\mathrm{Sc}}(z_{1}-z_{2}).\label{eq:1d_he_model}
\end{equation}
We solve the corresponding 
Schrödinger equation using this Hamiltonian to get the wave function
$\Psi(z_{1},z_{2},t)$. Let us remind that the latter
is a symmetric function in the two spatial variables. For the time-evolution,
we set the initial state as the field-free ground state at $t=0$
(using $\mathcal{E}_{z}(t\leq0)=0$). We handle the numerical time
evolution and the imaginary time-propagation (for obtaining the ground
state energy) with a split-step finite difference method \cite{wang2005nltdsedifferences,majorosi2016tdsesolve,majorosi2018densitymodel}.

In this two dimensional formalism we calculate the physical quantities
that can be derived from the following form of the reduced density
\begin{equation}
\varrho_{z}(z,t)=2\int\left|\Psi(z,z_{2},t)\right|^{2}\mathrm{d}z_{2}\label{eq:2d_density_z}
\end{equation}
from which the mean values and the root-mean-square deviations of
the spatial coordinate follow straightforwardly. We also make an approximate
formula of 
\begin{equation}
g(t)=1-\frac{1}{2}\left\vert \int\sqrt{\varrho_{z}(z,0)}\sqrt{\varrho_{z}(z,t)}\mathrm{d}z\right\vert ^{2}\label{eq:proj2_ground_loss}
\end{equation}
that we call ground state population loss per electron orbital. This
is directly comparable to the ground state population loss of the
3D Hartree-Fock formalism.

The model potential $V_{0,\mathrm{Sc}}(z)$ implicitly
depends on the fitting parameter $\alpha^{2}$, which is determined
by setting the initial two-electron energy. We use two types of parametrization.
We call the first ``ab initio”\ parametrization, in which we use
the single-electron model parameters of \eqref{eq:1d_params_soft} with
$Z=2$, corresponding to $Z^{\ast}=1$, and $\alpha^{2}=0.0625.$
For the ground state energy of this 1D model system we have obtained
$E_{0}=-3.02$ without additional parameter fitting. This is to be
compared with the ground state energy of the real helium: $-2.903$
\cite{drake1999he}, indicating an error of about $3.4\%$ in the
ground state energy of this ``ab initio”\ 1D model.

In the second parametrization of $V_{0,\mathrm{Sc}}(z)$, the $\alpha^{2}$
was modified to reproduce the 1D ground state energy $E_{0}=-2.903$
(i.e. it matches that of the real helium), with $Z^{\ast}=1$, and
$\alpha^{2}=0.0694$. Since the 3D\ Hartree-Fock method yields a
ground state energy $E_{0}=-2.860$, 
the corresponding 3D reference simulations 
have also been modified to match the accurate ground state energy
of $-2.903$. (To this end, the Hartree-potential was multiplied by
a factor that is slightly different from 1.)

\subsection{1D hydrogen molecule model\label{subsec:model_2d_h2}}

Finally, we create the 1D model of the hydrogen molecule in a similar way to the case of the He atom.
We use again a wave function of two variables $\Psi(z_{1},z_{2},t)$
and we  replace 
the true Coulomb potentials by our improved 1D soft-core potentials \eqref{eq:1d_pot_soft}
and \eqref{eq:1d_soft_pot_mol},
which gives the following Hamiltonian for the 1D model hydrogen molecule: 
\begin{equation}
H_{\mathrm{H_{2}}}^{\mathrm{1D}}=\sum_{k=1}^{2}\left[T_{z_{k}}+V_{0,\mathrm{Sc}}^{(\mathrm{M})}(z_{k};d)+z_{k}\mathcal{E}_{z}(t)\right]-\frac{1}{2}V_{0,\mathrm{Sc}}(z_{1}-z_{2}).\label{eq:1d_h2_model}
\end{equation}
Note that this Hamiltonian reduces to the one used for the helium
atom in Eq. \eqref{eq:1d_he_model} for the limiting value of the
internuclear distance $d=0.$ 
The 1D potentials $ V_{0,\mathrm{Sc}}^{(\mathrm{M})}$ and $V_{0,\mathrm{Sc}} $
depend implicitly on the parameters
$Z^{\ast}$ and $\alpha^{2}$ but we use identical parametrization for these potentials. 
We are going to test the case when $Z^{\ast}=0.5$
which equals to that of the hydrogen atom, and alternatively the case
when the asymptotics determined by $Z^{\ast}$ agree with the one
derived from the 1D hydrogen molecular ion model.

Regarding the energy values, Doma \cite{doma2016h2vmc} gave $-1.173427$
a.u. for the dissociation energy of the real 3D hydrogen molecule
with $d=1.4$, which means $-1.88729$ a.u. electronic ground state energy. Our reference
Hartree-Fock computation gives $E_{0}=-1.848$ a.u.  ground state energy
which means a relative error of about $2.1\%$ in our 3D reference
scheme. For consistency, the ground state energy of the 1D model system
is adjusted to this Hartree-Fock energy, by setting $\alpha{}^{2}$
to the proper value, as to be explained in section \ref{subsec:resultsH2mol}.

\section{Results and comparison of the 1D and 3D simulations\label{sec:results}}

In this section, we present and compare the results of strong-field
simulations of the 1D models of the previous sections to the results
acquired from the 3D reference models.

\begin{figure}[h]
\includegraphics[width=1\columnwidth]{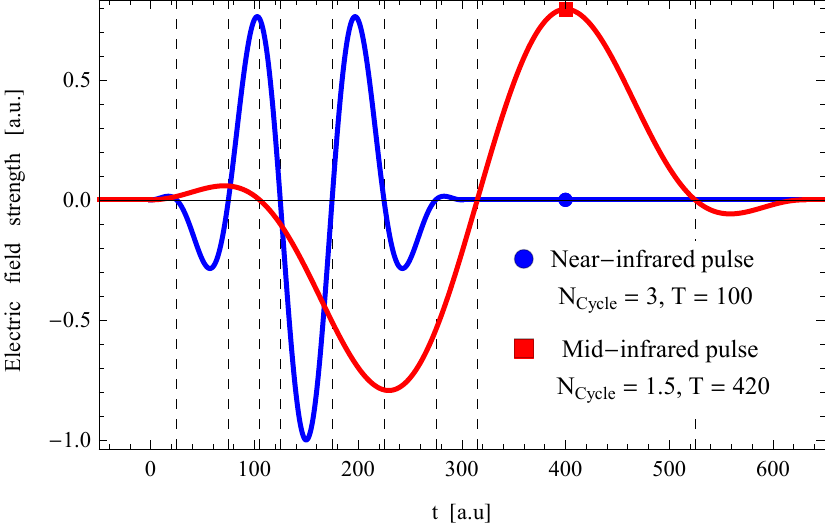}
\protect\caption{Pulse shape of the near-infrared (in blue) and mid-infrared (in red) laser pulse, 
with the indicated parameters corresponding to Eq. (\ref{eq:sim_sinpulse3_E}).}

\label{fig:pulseshape}
\end{figure}

In these simulations, we model the linearly polarized few-cycle laser
pulse with a sine-squared envelope function. The corresponding time-dependent
electric field has non-zero values only in the interval $0\leq t\leq N_{\text{Cycle}}T$
according to the formula: 
\begin{equation}
\mathcal{E}_{z}(t)=F\cdot\sin^{2}\left(\frac{\pi t}{N_{\text{Cycle}}T}\right)\cos\left(\frac{2\pi t}{T}\right),\label{eq:sim_sinpulse3_E}
\end{equation}
where $T$ is the period of the carrier wave, $F$ is the peak electric
field strength and $N_{\text{Cycle}}$ is the number of cycles under
the envelope function. 
In subsections \ref{subsec:results_H2+}-\ref{subsec:resultsH2mol}, we model a near-infrared laser pulse by setting
 $T=100$, corresponding to a ca. $725\,\mathrm{nm}$ carrier wavelength, and $N_{\text{Cycle}}=3$
 which gives a pulse with its main peak at its center.
In order to simulate the effect of a mid-infrared laser pulse in subsection \ref{subsec:resultsLONG}, we set $T=420$, corresponding to a  carrier wavelength of  $3045\,\mathrm{nm}$, and  $N_{\text{Cycle}}=1.5$
which gives a pulse 
with a zero at its center.
We plot these pulse shapes in Fig. \ref{fig:pulseshape}.
From Fig. \ref{fig:pulseshape}
on, the vertical dashed lines denote the zero crossings of the respective
$\mathcal{E}_{z}(t)$ electric field.

In our calculations, we set the typical step sizes as $\Delta z=0.2$
and $\Delta t=0.01$ since these are sufficient for the numerical
errors to be within line thickness. We use box boundary conditions
and we set the size of the box to be sufficiently large so that the
reflections are kept below $10^{-8}$ atomic units in the wave function.

The results belonging to the correct reference 3D simulation of a
given system are plotted in blue and labeled as ``3D-reference''.
We also plot dashed blue overlays for these reference curves for clarity.

\subsection{Hydrogen molecular ion driven by a near-infrared laser pulse \label{subsec:results_H2+} }

\begin{figure*}
\begin{raggedright} \hspace{4.5cm}(a)\hspace{8.7cm}(b)

\end{raggedright}

\includegraphics[width=1\columnwidth]{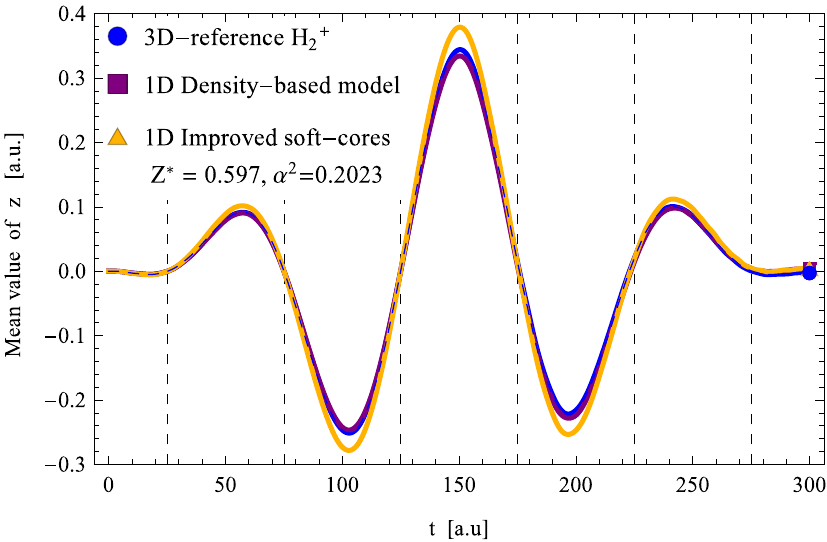}\hspace{0.5cm}\includegraphics[width=1\columnwidth]{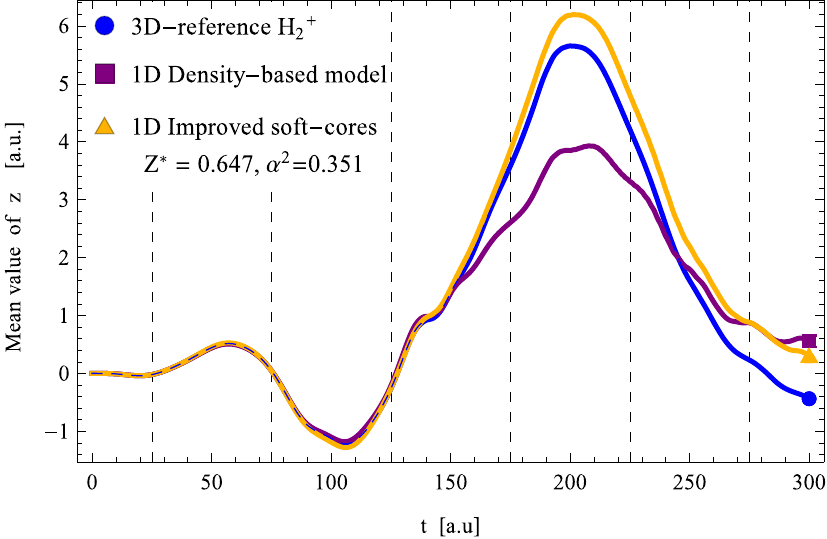}

\begin{raggedright} \hspace{4.5cm}(c)\hspace{8.7cm}(d)

\end{raggedright}

\includegraphics[width=1\columnwidth]{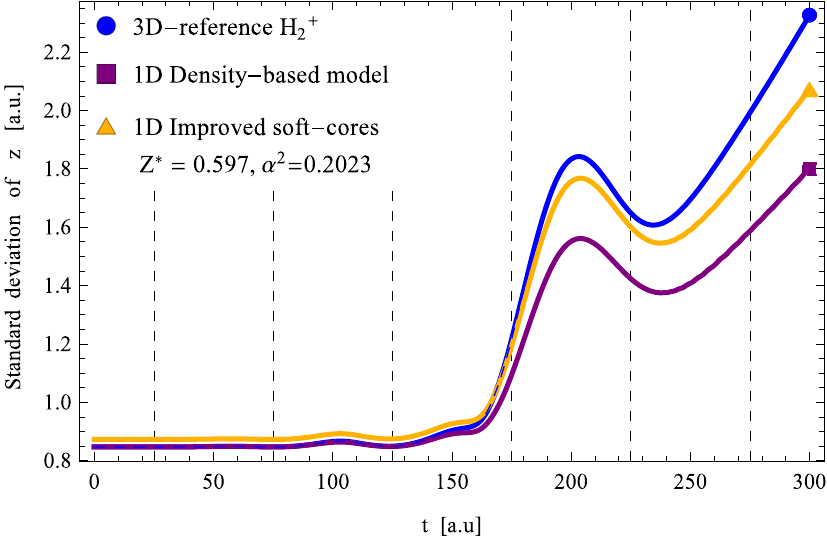}\hspace{0.5cm}\includegraphics[width=1\columnwidth]{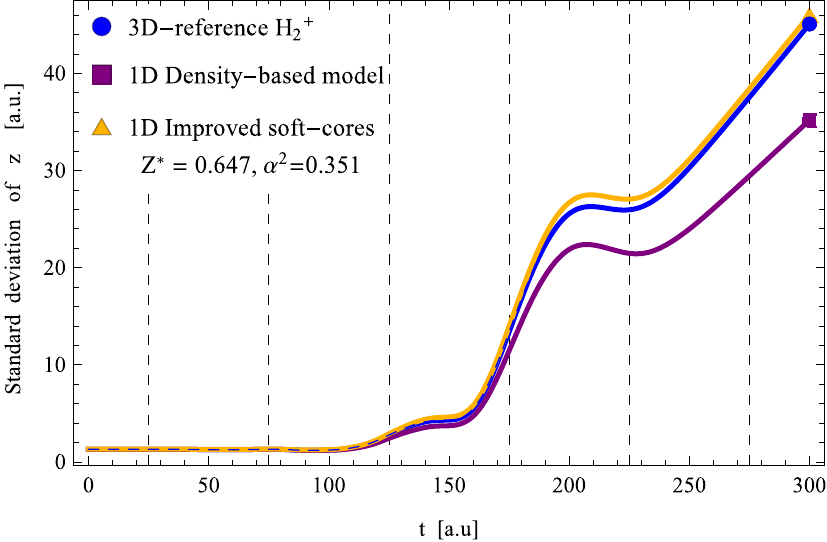}

\begin{raggedright} \hspace{4.5cm}(e)\hspace{8.7cm}(f)

\end{raggedright}

\includegraphics[width=1\columnwidth]{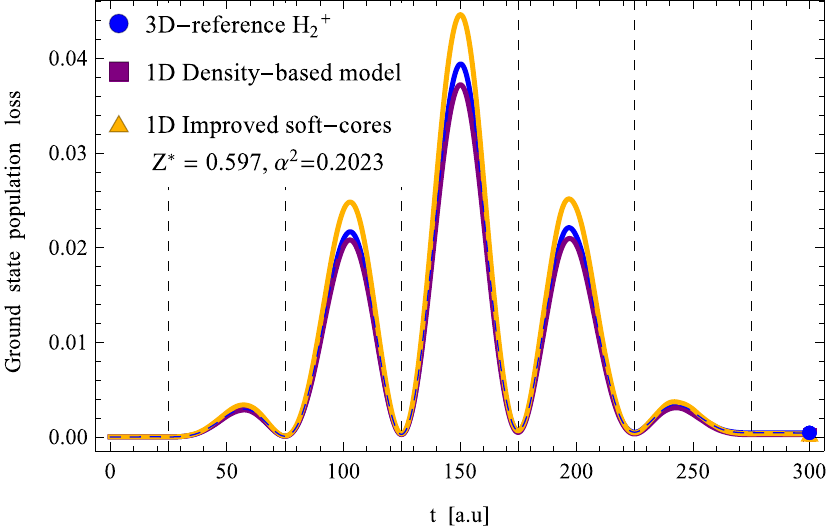}\hspace{0.5cm}\includegraphics[width=1\columnwidth]{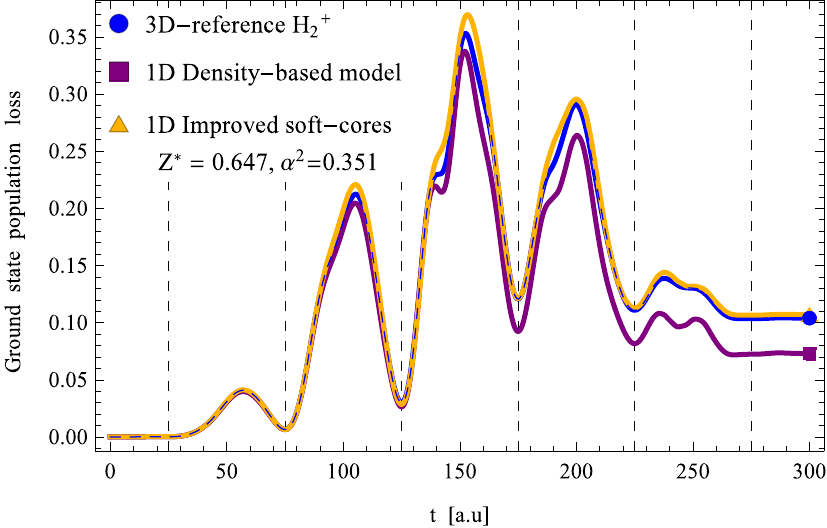}

\protect\protect\protect\caption{Results for two static hydrogen molecular ions with  intermolecular
distances of $d=1.4$ (left panels) and $d=2.6$ (right panels),
driven by the    {near-infrared} laser pulse with $F=0.15$. We show
the time-dependence of the mean values $\left\langle z\right\rangle (t)$
in (a)-(b) and the standard deviations $\sigma_{z}(t)$ in (c)-(d) and the
ground state population losses $g(t)$ in (e)-(f). Results of the corresponding
3D simulations are plotted in blue.}

\label{fig:SinPulse3_D26_H2+} 
\end{figure*}

In this section we present our results for the hydrogen molecular
ion with two selected  external field amplitudes $F$ and three selected
 molecular distance parameters $d$. We have investigated the following model potentials,
the results of which are plotted on each of the figures belonging
to this section (from Fig. \ref{fig:SinPulse3_D20_H2+} to Fig. \ref{fig:SinPulse3_D26_H2+}):
the density-based molecular 1D model potential \eqref{eq:1d_density_pot_mol}
(in purple), and the 1D soft-core model \eqref{eq:1d_soft_pot_mol}
using the correct asymptotes (in gold). In these figures we plot the
time-dependence of the following physical quantities: the mean value
$\left\langle z\right\rangle (t)$, the standard root-mean-square
deviation $\sigma_{z}(t)$, and the ground state population loss $g(t)$,
as determined by the time-dependent wave function.

Let us begin the discussion with a hydrogen molecular ion frozen in
its equilibrium distance ($d=2.0$). From time-independent calculations
of the reduced ground-state density, we inferred that the correct
1D density-based model does not behave like the composition of two
1D density-based atomic binding potentials having Coulomb asymptotes
with $Z^{*}=0.5$. Instead, in this case, these long range 1D Coulomb
asymptotes have the value of around $Z^{*}=0.625$, that we obtained
by numerical calculations based on \eqref{eq:1d_density_pot_mol}.
This means that for this separation distance, the Coulomb cores exhibit
a screening-like effect in the 1D model. Accordingly, we parametrized
the 1D improved soft-core model using $d=2.0$ with $Z^{*}=0.625$
and $\alpha^{2}=0.277$.

In Fig. \ref{fig:SinPulse3_D20_H2+} we can see the time-evolution
results for this model system under the influence of the external
fields with $F=0.1$ (left panels) and $F=0.15$ (right panels),
corresponding to a peak intensity of $3.51\times10^{14}\,\mathrm{W/cm}^2$ and $7.89\times10^{14}\,\mathrm{W/cm}^2$, respectively.
These
results have comparable accuracy to that of our previous calculations
in Ref. \cite{majorosi2018densitymodel} using single-electron atom
models (although we use larger $F$ values here). In the case of $F=0.1$,
hardly any ground state population loss occurs after the laser pulse,
even though temporarily it raises to relatively larger values, in
sync with the laser field. This behavior is quantitatively correctly
predicted by our 1D models. The curves of $\sigma_{z}(t)$  show that
the 1D improved soft-core model has somewhat lower accuracy for the
 1D ground state ($\sigma_{z}(t)$ has a somewhat larger initial value
 in the soft-core case  compared to 3D case), but its laser-field
induced dynamics  follows better the 3D results in overall. This latter
is even more pronounced for the stronger field value $F=0.15$. The
curves of the mean value $\left\langle z\right\rangle (t)$ also show
good quantitative agreement.

Although these 1D models qualitatively reproduce even the finer details
of the 3D ionization dynamics of $\mathrm{H}_{2}^{+}$, we should
note that the 1D models yield high-order harmonics with larger amplitudes
than the 3D results. This is mainly caused by the fact that the interference
effects in the 1D density are stronger than in the reduced 3D density
(since these latter are suppressed by the spatial integration along
the $\rho$ coordinate). This feature of the HHG spectra was successfully
accounted for by the scaling function \eqref{eq:1d_power_scale} for
a hydrogen atom \cite{majorosi2018densitymodel}, which turned out
to be suitable also for the present case of $\mathrm{H}_{2}^{+}$.

Next, let us discuss hydrogen molecular ion configurations where the
parameter $d$ is different from its equilibrium value $d=2.0$. We
chose for the case of closer nuclei $d=1.4$, which may be regarded
as an instantaneously ionized static hydrogen molecule. Based on the
calculation of the 1D density-based potential \eqref{eq:1d_density_pot_mol},
we computed the parameter of the effective Coulomb asymptotes as $Z^{*}=0.597$,
having a ground state energy of $E_{0}=-1.284$. Accordingly, the
fitting parameter has to be $\alpha^{2}=0.2023$. For the case of
nuclei with a larger distance, we chose the value $d=2.6$, which
is closer to limit of the molecular dissociation. From the calculation
of the 1D density based potential \eqref{eq:1d_density_pot_mol} we
get its ground state energy as $E_{0}=-0.975$, and the parameters
of the Coulomb asymptotes as $Z^{*}=0.647$ and $\alpha^{2}=0.351$.
Note that the dependence of these parameters on $d$ is not negligible,
and the parameter $Z^{*}$ appears to be increasing as $d$ is increasing.
However, for nuclei very far apart, the asymptotic value of $Z^{*}=0.5$
should hold to describe a 1D H-atom and a proton. For the other limiting
case, $d=0$, we get the 1D $\mathrm{He}^{+}$ with the value of $Z^{*}=0.5$
again.

In Fig. \ref{fig:SinPulse3_D26_H2+} we plot the results for $d=1.4$
(left panels) and $d=2.6$ (right panels) using the external electric
field with $F=0.15$. We can see that the induced dynamics with $d=1.4$
is similar in behavior to the case of $d=2.0$ using  $F=0.1$ (see Fig. \ref{fig:SinPulse3_D20_H2+}). In
the case of $d=2.6$, the ground state population loss is much larger
 than in the previous cases with the same $F$ value. Most importantly,
both of these 1D models reproduce even the finer details of the curves
of $g(t)$, $\sigma_{z}(t)$, $\left\langle z\right\rangle (t)$ of
the true 3D dynamics.

We can also observe that the accuracy of the results improves as the
value of $d$ increases, especially for the 1D improved soft-core
potential, 
which is mainly due to the fact that the probability concentrated
between the nuclei decreases with increasing $d$. This increasing
accuracy suggests that these 1D potentials are suitable also to model
strong field processes leading to molecular dissociation.

\subsection{Helium atom driven by a near-infrared laser pulse,  high-order-harmonic spectra \label{subsec:resultsHe}}

\begin{figure*}
\begin{raggedright} \hspace{4.5cm}(a)\hspace{8.7cm}(b)

\end{raggedright}

\includegraphics[width=1\columnwidth]{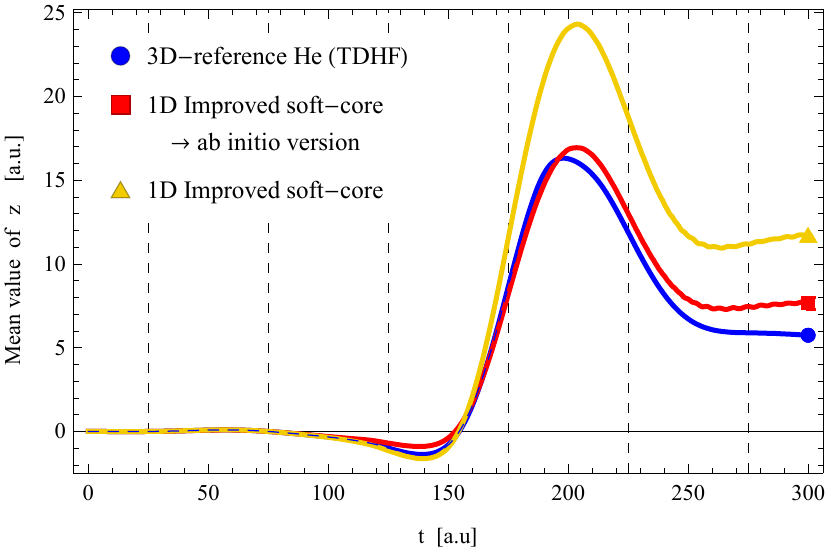}\hspace{0.5cm}\includegraphics[width=1\columnwidth]{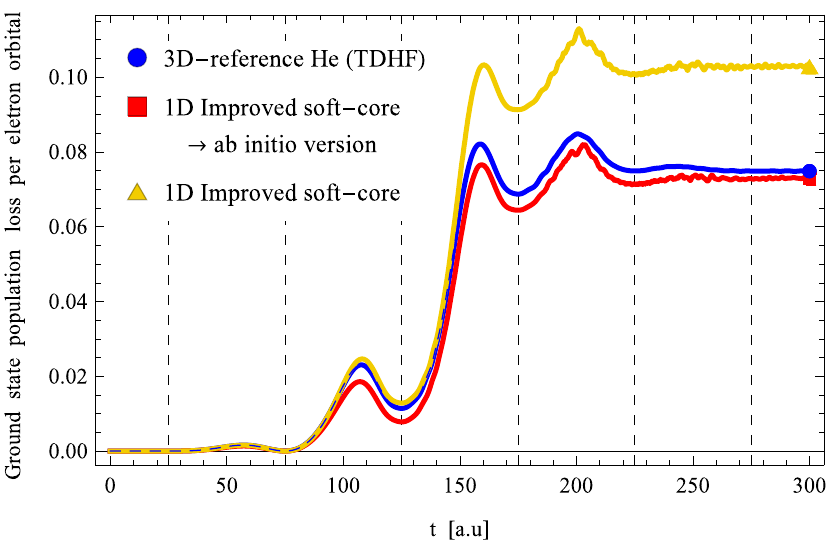}

\protect\protect\protect\caption{The time-dependence of the mean values
$\left\langle z\right\rangle (t)$ in (a) and the ground state population
losses $g(t)$ in (b) for a helium atom driven by the    {near-infrared} laser pulse with
$F=0.25$.}

\label{fig:SinPulse3_He_F0.25} 
\end{figure*}

\begin{figure*}
\begin{raggedright} \hspace{4.5cm}(a)\hspace{8.7cm}(b)

\end{raggedright}

\includegraphics[width=1\columnwidth]{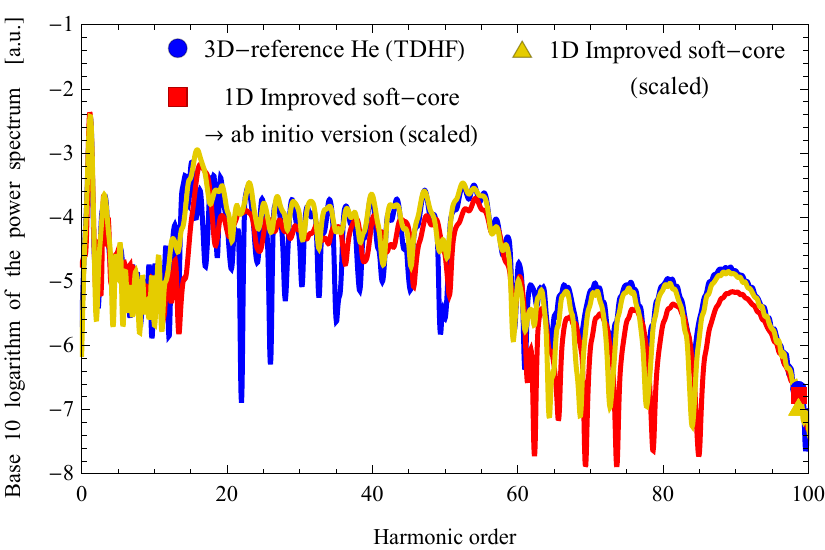}\hspace{0.5cm}\includegraphics[width=1\columnwidth]{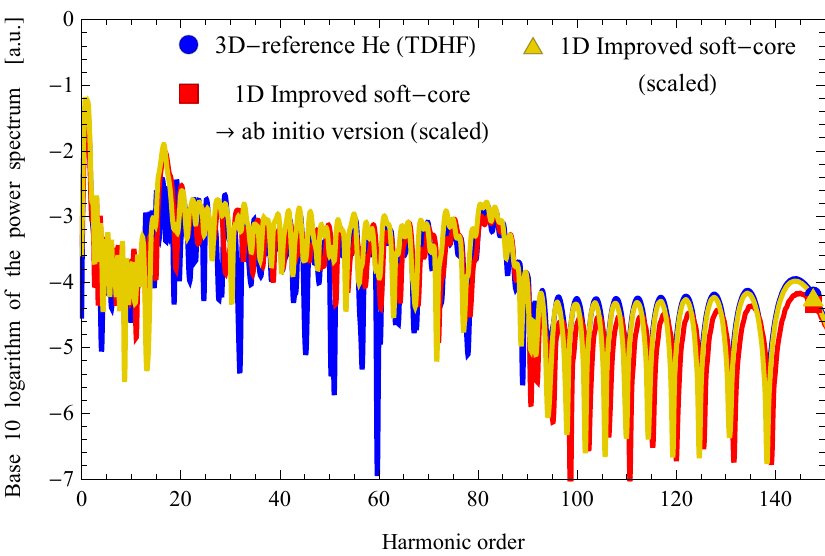}

\begin{raggedright} \hspace{8.7cm}(c)

\end{raggedright}

\includegraphics[width=2.07\columnwidth]{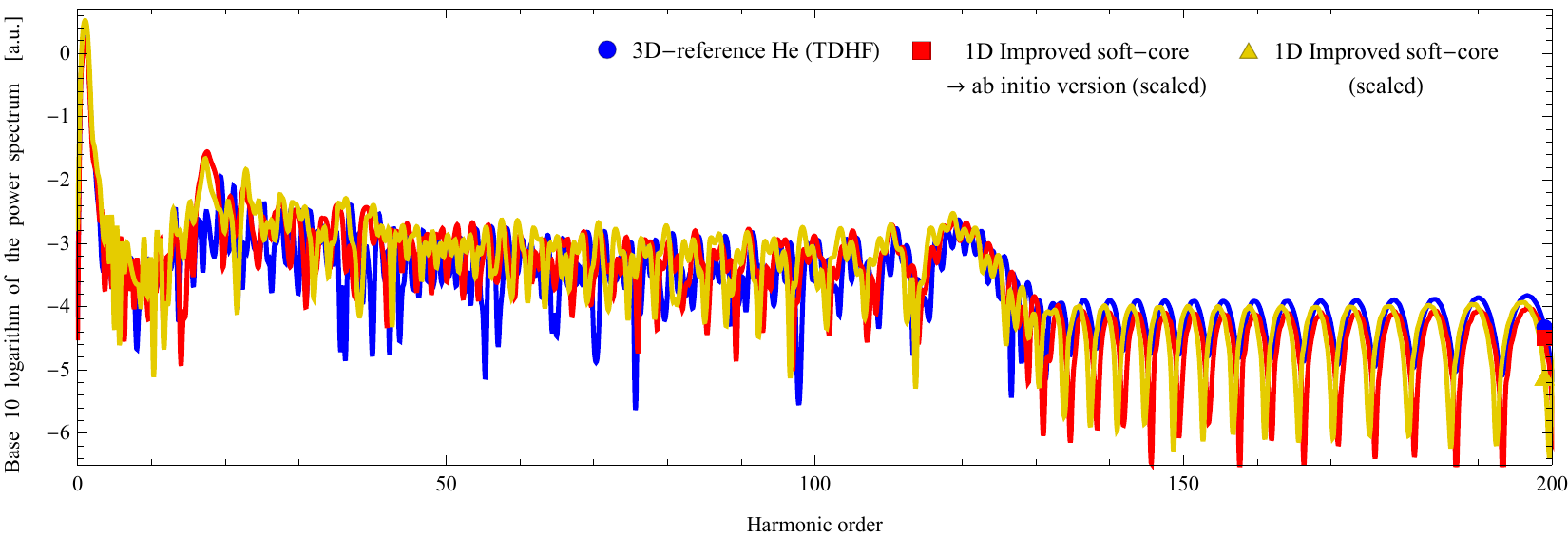}

\protect\protect\protect\caption{The scaled power spectra versus the harmonic
order of the emitted radiation for a helium atom driven by the    {near-infrared} laser pulse with $F=0.15$ in (a), $F=0.20$ in (b), and $F=0.25$ in (c). Results
of the corresponding 3D simulations are plotted in blue. The scaling
function      {of Eq. (\ref{eq:2d_power_scale_he}) was applied to all of the 1D results of this figure}.}

\label{fig:SinPulse3_He_power} 
\end{figure*}

In the following, we present our results regarding the 1D model helium
atom based on Sec. \ref{subsec:model_2d_helium} under the influence
of an external laser pulse, for selected values of $F$. In Figs.
\ref{fig:SinPulse3_He}-\ref{fig:SinPulse3_He_power}, we plot results
of the following 1D model soft-core potentials: $(i)$ the 1D improved
soft-core potentials with ``ab initio'' parametrization ($Z^{*}=1$,
$\alpha^{2}=0.0625$) from \eqref{eq:1d_params_soft} in red, $(ii)$
the 1D improved soft-core potentials with energy fitting ($Z^{*}=1$,
$\alpha^{2}=0.0694$) in gold, and $(iii)$ the 1D usual soft-core
potentials \cite{bauer1997tdse1dhemodel} using the normal Coulomb
asymptotes and with the same form of energy fitting as the previous
case ($Z^{*}=2$, $\alpha^{2}=0.5474$) in green. Except for $(i)$,
the time evolution of the 1D two-electron model systems starts from
a ground state that has the same energy as the reference time-dependent
Hartree-Fock simulation.

We show in Fig. \ref{fig:SinPulse3_He} the time-dependence of the
mean values $\left\langle z\right\rangle (t)$, the standard root-mean-square
deviations $\sigma_{z}(t)$ and the ground state population losses
per electron orbital $g(t)$, with $F=0.15$ (left panels) and $F=0.2$
(right panels),
corresponding to a peak intensity of   $7.89\times10^{14}\,\mathrm{W/cm}^2$ and $1.40\times10^{15}\,\mathrm{W/cm}^2$, respectively.
For $F=0.15$, we can see that the dynamics induced
by the laser field are weak, which is  true both for the 3D reference
and the 1D improved soft-core models, while the results of the usual
1D model are significantly off. Our ``ab initio'' parametrization
$(i)$ of the 1D improved soft-core Coulomb forms is still quantitatively
acceptable, and corresponds to a stronger bound than the real helium
atom. This latter seems to be in agreement with the fact that it has
a lower bound state energy by $0.1$ a.u. If we now look at the
results corresponding to $F=0.2$, we can see that the results of
our improved models are similar, albeit slightly off with stronger
ionization. Based on the similarity to the results of $\left\langle z\right\rangle (t)$
and $\sigma_{z}(t)$, we also note that the  quantity $g(t)$ defined
in \eqref{eq:proj2_ground_loss} indeed bears the same meaning as
in \eqref{eq:proj3_ground_loss} calculated from the Hartree-Fock
orbital.

We also plot in Fig. \ref{fig:SinPulse3_He_F0.25} the $\left\langle z\right\rangle (t)$
and $g(t)$ results for $F=0.25$. We can see now that the 1D results
are relatively up shifted compared to the previous figures in such
a way that now the curves of 1D ``ab initio'' parametrization much
better match the 3D reference. Since such an effect does not occur
in the 1D models for the hydrogen molecular ions (or hydrogen atoms)
under similar conditions, this may indicate the inaccuracy of our
approximations at such high intensities, which presumably affects
also the Hartree-Fock calculations.

Let us next investigate the high-order harmonic response of the particular
1D model helium. The accurate computation of the high-order harmonic
spectrum is especially important in strong-field physics: its well-known
characteristic features \cite{McPherson_JOSAB_1987_HHG,Ferray_JPhysB_1988_HHG,Harris_OptCom_1993_HHG_Atto,Krausz_RevModPhys_2009_Attosecond_physics,gombkoto2016quantoptichhg}
represents the highly nonlinear atomic response to the strong-field
excitation, and its suitable phase relations enable the generation
of attosecond pulses of XUV radiation \cite{Farkas_PhysLettA_1992_AttoPulse,Paul_Science_2001_Atto_pulsetrain,hentschel2001attosecond,drescher2002attosecond,kienberger2002attosecond,carrera2006attoxuv,sansone2006attosecondisolated}.
In accordance with Ref. \cite{majorosi2018densitymodel} we expect
that the structure 
 of the spectra based on the 1D and the 3D simulations is similar, 
 but the amplitudes are 
 larger
in the 1D results. The reason for the latter is that
rescattering
on the ion-core is a much stronger effect  in a 1D dynamics than in a 3D  dynamics (amplifying the high-frequency oscillations in the 1D results), 
and  on the other hand, 
the integration
over the transverse directions decreases the effect of the oscillations of the 3D wave function on the  mean values, like  $\left\langle z\right\rangle (t)$.
(These also cause the small oscillations are visible only on the 1D curves, but not on the corresponding 3D curves,  of some of the plots in Figs.  \ref{fig:SinPulse3_He} and \ref{fig:SinPulse3_He_F0.25}).
This feature of the 1D models 
can be handled again by introducing a frequency dependent
scaling function. In Fig. \ref{fig:SinPulse3_He_power} we can see
the scaled $p(f)/s(f)$ power spectrum of the second derivative of
$\left\langle z\right\rangle (t)$, where we applied the following
scaling function: 
\begin{equation}
s(f)=\min\left(1+0.01\left(100f-1\right)^{3},1+2.7\left|100f-1\right|\right),\label{eq:2d_power_scale_he}
\end{equation}
which we obtained by fitting its parameters to reproduce spectrum
for the case $F=0.15$. In Fig. \ref{fig:SinPulse3_He_power} (a)
we can see that the structure of the scaled spectra is indeed a good
match compared to the reference Hartree-Fock results. The spectrum for
the ``ab initio'' case  has a similar structure, and it is shown
with the same scaling function. If we compare \eqref{eq:2d_power_scale_he}
to \eqref{eq:1d_power_scale} used for a single electron atom, we
 see that the 1D helium atom needs a stronger scaling (by about $2.7$
times) especially for the higher harmonics. What makes this scaled
spectrum interesting, that it works for different configuration of
electric fields: in Fig. \ref{fig:SinPulse3_He_power} (b) and \ref{fig:SinPulse3_He_power}
(c) we applied the same scaling. For the larger field value of $F=0.2$
the improved soft-core results still replicate the 3D spectra really
well. (However, there seems to be some deviation near the 20th harmonic.)
For $F=0.25$, we can see that the spectrum is scaled properly, but
the positions of the harmonics are slightly shifted from the 3D reference.
Here, the ``ab initio'' model better describes the structure of
the spectra, which is connected to the closer matching to the 3D reference
of the other quantities of Fig. \ref{fig:SinPulse3_He_F0.25}.

Finally we note that the match of the spectral phase is also very
good, especially in the higher frequency range, which is of fundamental
importance for the generation of isolated attosecond pulses.

In overall we can say that, the 1D improved soft-core models replicate
the strong-field response of a real helium atom comparing to the time-dependent
Hartree-Fock approximation up to around $F=0.2$. This includes the
low frequency response of the mean motion on the level of the reduced
density. The low dimensionality however, gives a much larger high-order-harmonic
amplitude, which can be converted to the corresponding 3D spectra
using one scaling function. The improved models discussed here appear
to be quantitatively correct 1D models of the helium atom.

The merits of these 1D model potentials may also pave the way to simulate
properties of a dilute medium, like an atomic gas-jet, used in actual
strong field or attosecond physics experiments with He \cite{shafir2012tunneling, chen2012heliumabsorb, stooss2018strongfieldresponse}. Previously, this was done \cite{tosa2016propagation,heyl2016nonlineargases}
by calculating Lewenstein's integral \cite{lewenstein1994hhgtheory},
but using the density-based model potentials and integrating the low
dimensional TDSE has now become also an option \cite{Berman2019intense_pulse_prop_gas}.

\subsection{Hydrogen molecule driven by a near-infrared laser pulse \label{subsec:resultsH2mol}}

\begin{figure*}
\begin{raggedright} \hspace{4.5cm}(a)\hspace{8.7cm}(b)

\end{raggedright}

\includegraphics[width=1\columnwidth]{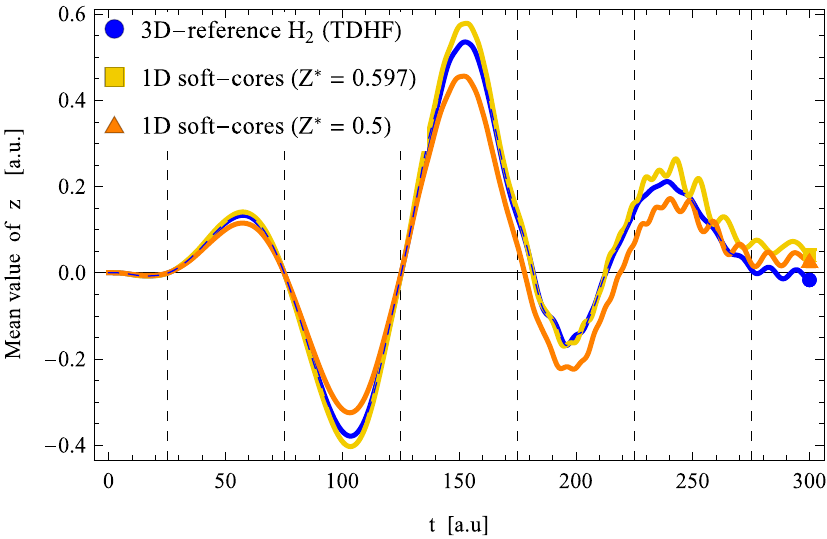}\hspace{0.5cm}\includegraphics[width=1\columnwidth]{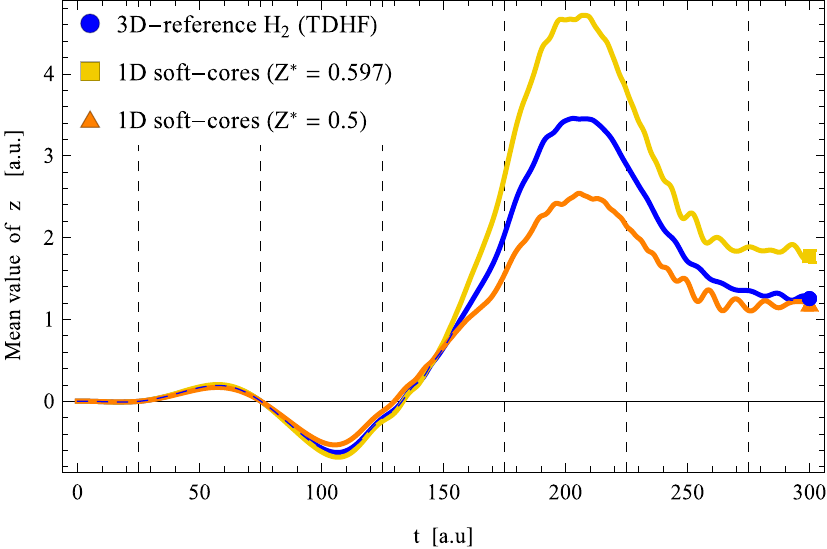}

\begin{raggedright} \hspace{4.5cm}(c)\hspace{8.7cm}(d)

\end{raggedright}

\includegraphics[width=1\columnwidth]{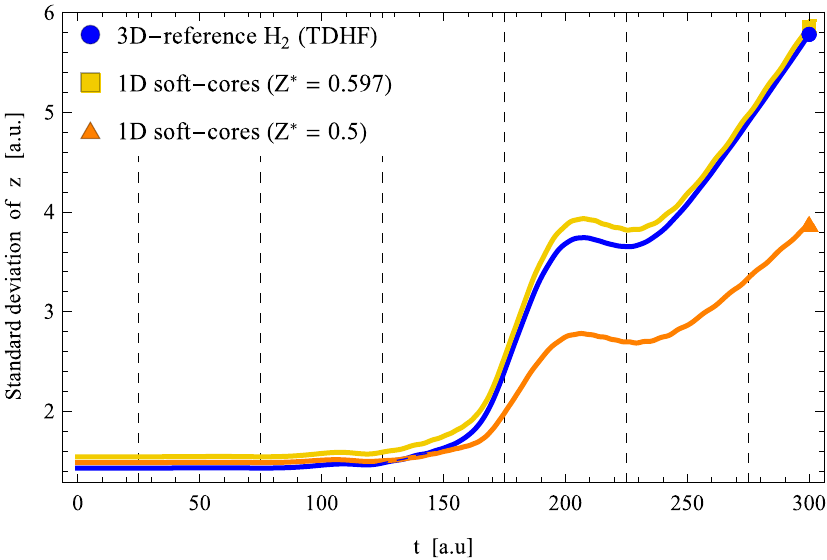}\hspace{0.5cm}\includegraphics[width=1\columnwidth]{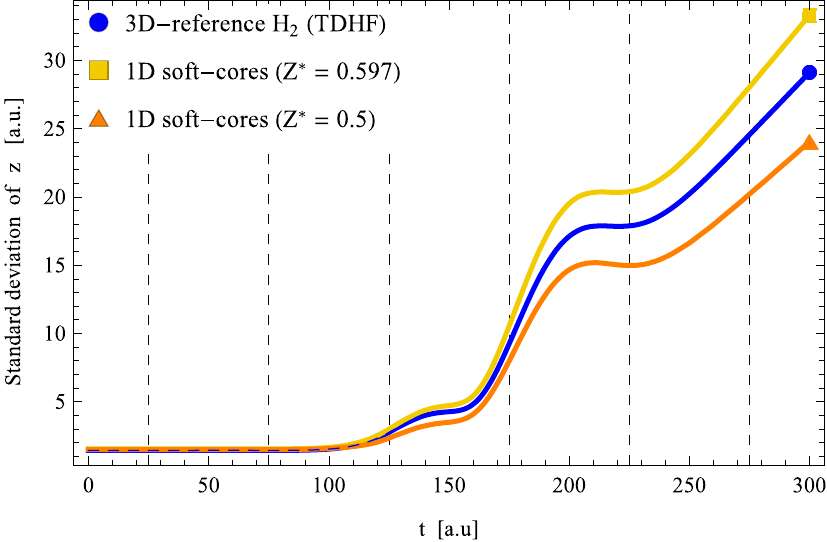}

\begin{raggedright} \hspace{4.5cm}(e)\hspace{8.7cm}(f)

\end{raggedright}

\includegraphics[width=1\columnwidth]{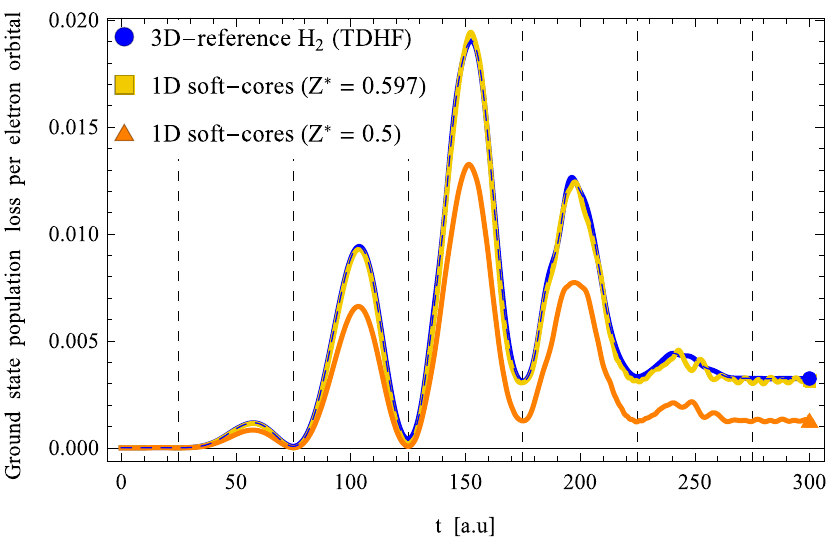}\hspace{0.5cm}\includegraphics[width=1\columnwidth]{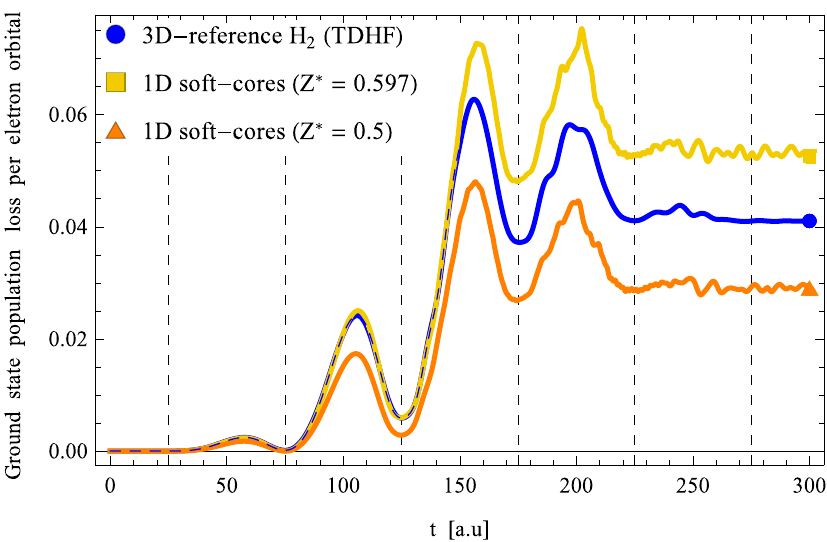}

\protect\protect\protect\caption{Results for a static hydrogen molecule with $d=1.4$, 
 driven by the    {near-infrared} laser pulse with $F=0.07$ (left panels) and $F=0.1$ (right panels). We plot the time-dependence of the mean values $\left\langle z\right\rangle (t)$ in (a)-(b) and
the standard deviations $\sigma_{z}(t)$ in (c)-(d) and the ground state
population losses $g(t)$ in (e)-(f). Results
of the corresponding 3D simulations are plotted in blue.}

\label{fig:SinPulse3_H2} 
\end{figure*}

We present now the results for the 1D model hydrogen molecule as introduced
in Sec. \ref{subsec:model_2d_h2}. We apply the same two-electron
formalism as for the helium atom, and we compare again the results
with the corresponding 3D Hartree-Fock simulation as the reference.
We also make use of our results given in Sec. \ref{subsec:results_H2+}
regarding the dependence of the Coulomb asymptotes on the intermolecular
distance $d$ in the case of the hydrogen molecular ion model. Now
the presence of the extra electron poses the question, whether it
affects the values of the Coulomb asymptotes. To get the answer, we
have considered two different parametrizations of the improved soft-core
model $V_{0,{\rm Sc}}^{({\rm M})}$ in Eq. \eqref{eq:1d_h2_model}:
$(i)$ the model using the corresponding \emph{hydrogen molecular
ion} asymptotes ($d=1.4$, $Z^{\ast}=0.597$, $\alpha^{2}=0.235$)
and $(ii)$ soft-core models using \emph{atomic} asymptotes ($d=1.4$,
$Z^{\ast}=0.5$, $\alpha^{2}=0.112$). The $\alpha^{2}$ parameters
were determined by fitting the ground state energies, so that the
time evolution of the 1D two-electron model systems started from a
ground state that had the same energy value as the reference (3D)
time-dependent Hartree-Fock simulation. It turned out that a rather
good agreement could be obtained with the asymptotes from the H$_{2}^{+}$
model, i.e. from the choice $(i)$, especially for lower excitation
amplitudes. This is demonstrated in Fig. \ref{fig:SinPulse3_H2} \ where
the results are shown for two specific values: $F=0.07$ (left panels)
and $F=0.1$ (right panels), in yellow for model $(i).$ For comparison
the functions obtained with model $(ii)$ are also shown in orange.

The two particular $F$ values for which we show the results for H$_{2}$
were chosen because the induced ionization response is similar in
magnitude to that of the He atom with the amplitudes $F=0.15$ and
$F=0.2$ shown in Fig. \ref{fig:SinPulse3_He}. If we compare the
left panels of of Fig. \ref{fig:SinPulse3_H2} and Fig. \ref{fig:SinPulse3_He}, we can see that the yellow
curves of the physical quantities $\left\langle z\right\rangle (t)$
and $\sigma_{z}(t)$, behave similarly and they show a good match
with the 3D reference results if the ionization is low. We also note
that the quantity $g(t)$ behaves in the same way, compared to the
reference. \ This agreement is a consequence of the picture that
considers the He atom as the $d=0$ limit of H$_{2},$ but in the
case of He, which has more strongly bound electrons, larger electric field amplitudes
are required to achieve a similar effect. In the right panels of the
respective figures, which show simulation results with higher intensities
we can see that for model $(i)$ an up-shift occurs, the size of which
in Fig. \ref{fig:SinPulse3_H2} is again consistent with the results
of Fig. \ref{fig:SinPulse3_He}. This value of up-shift may still
be quantitatively acceptable for a one dimensional molecular model,
especially because the reference Hartree-Fock method may certainly
become inaccurate at these larger field strengths. We also note, that
for higher intensities this up-shift becomes larger, causing that
already for $F=0.15$ the model $(ii)$ matches the 3D reference better (similar
to Fig. \ref{fig:SinPulse3_He_F0.25}, but not shown). From this we can conclude
that using the asymptote value $Z^{\ast}=0.597$ from the respective
static hydrogen molecular ion configuration behaves like the 1D improved
soft-core model of the helium calculations, i.e. it gives quantitatively
correct results for the physical quantities in a similar manner shown
here.

Overall it is impressive that using improved asymptotes with one dimensional
soft-core Coulomb potentials the physics of a 3D hydrogen molecule
becomes quantitatively reproducible by the corresponding 1D model system. From the
results we can say that indeed the correct model is based on the density
based potential, and its asymptotes are surely between the respective
hydrogen molecular ion asymptote with $Z^{*}=0.597$ and the hydrogen
atom asymptote with $Z^{*}=0.5$ in the tested peak electric field
strength range. For more accurate tests, more accurate 3D reference
simulation methods are necessary.

\subsection{$\mathrm{H}_2^+$ and He driven by a mid-infrared laser pulse \label{subsec:resultsLONG}}

Although the established laser technology for strong-field physics is mainly in the spectral range from ultraviolet  to near-infrared currently, 
the use of longer carrier wavelengths has several important advantages 
\cite{wolter2015strongfieldmidIR}, 
which makes the use of mid-infrared 
laser pulses a promising new research direction in strong-field physics 
\cite{wolter2014statesstrongmidIR,Kubel_Corkum_2017_PhysRevLett.119.183201_meas_1d-tdse,Quan_2017_PhysRevA.96.032511_meas,
Ortmann_Landsman_2017_PhysRevLett.119.053204_theo,Pullen_Biegert_2017_PhysRevA.96.033401_meas,Shao_2017,
Maurer_Keller_2018_PhysRevA.97.013404_meas-ellpol-Xe,Shaaran_Camus_Moshammer_2019_PhysRevA.99.023421_meas-Ar-Ne_theo-SFA,Ghimire_NatPhys_2019}.
However, the continuum electron wavepackets may travel much larger distances   with increasing wavelength in the case of gas targets, 
which rises the numerical demand of the corresponding 3D quantum simulations  sharply 
\cite{Thumm_2014_PhysRevA.89.063423_3d-tdse-midIR,Madsen_2016_PhysRevA.93.033420_3d-tdse-midIR-speedup}.
Thus, accurate 1D simulations are very important, therefore we present below a few promising results obtained with our 1D models for the
hydrogen molecular ion and the helium atom, driven by 
the mid-infrared laser pulse specified after Eq. (\ref{eq:sim_sinpulse3_E}). 
(Since the 1D He atom models show similar accuracy as the 1D $\mathrm{H}_2$ models do, 
we do not consider the latter here.)

In particular, we test our 1D models of $\mathrm{H}_2^+$
(with $d=2.0$) and He in the same manner as we did in Sec. \ref{subsec:results_H2+}
and \ref{subsec:resultsHe}, respectively, 
with some simulation parameters adjusted to the requirements due to the mid-infrared laser pulse. 
We perform
the simulations from 0 to 630 atomic time units, and we ensure that the
box size is large enough to keep the numerical errors within line thickness
by setting $z_{\min}=-2000$, $z_{\max}=2000$,
$\rho_{\min}=0$, $\rho_{\max}=1750$ for all
simulations in this subsection.
Note also that the mid-infrared pulse has a zero crossing
at its center (at $t=315$), and the two main peaks in opposite directions 
at $t=229.6$ and $t=400.4$ have a magnitude of ca. $0.795 F$.
 The anti-symmetry of this pulse with respect to its center helps to keep the electron's motion confined 
 which makes the 3D simulations less demanding and more accurate.

\begin{figure*}
\begin{raggedright} \hspace{4.5cm}(a)\hspace{8.7cm}(b)

\end{raggedright}

\includegraphics[width=1\columnwidth]{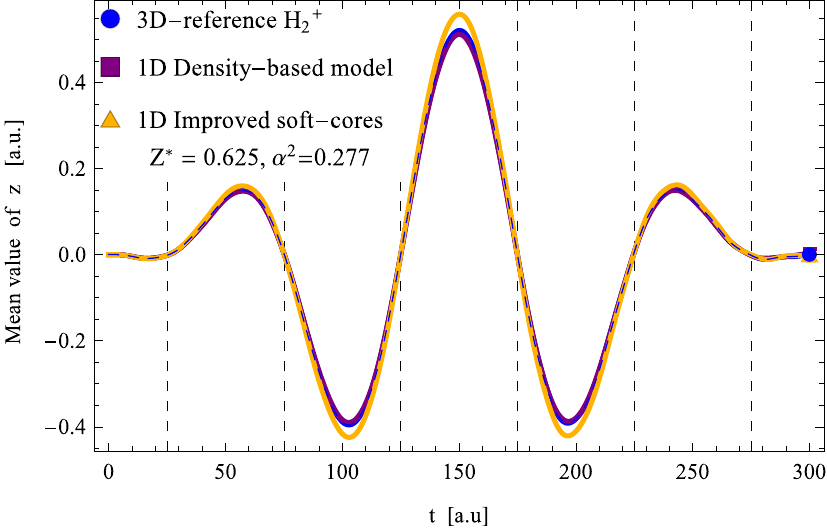}\hspace{0.5cm}\includegraphics[width=1\columnwidth]{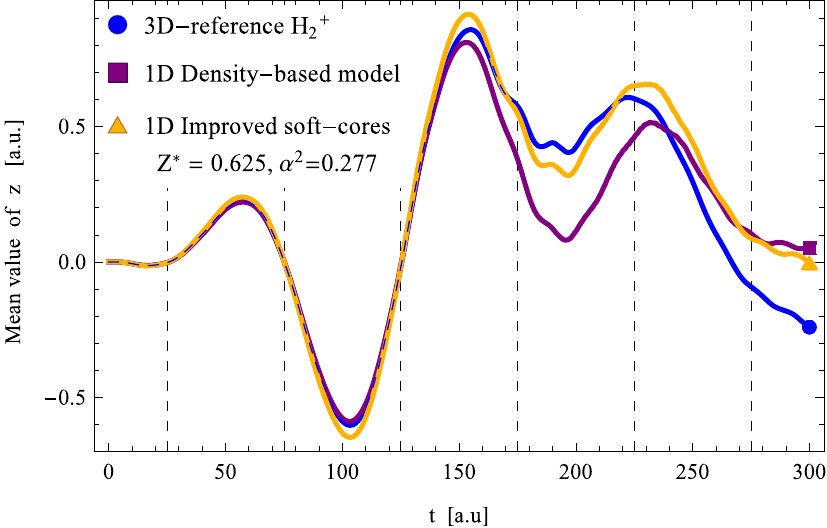}

\begin{raggedright} \hspace{4.5cm}(c)\hspace{8.7cm}(d)

\end{raggedright}

\includegraphics[width=1\columnwidth]{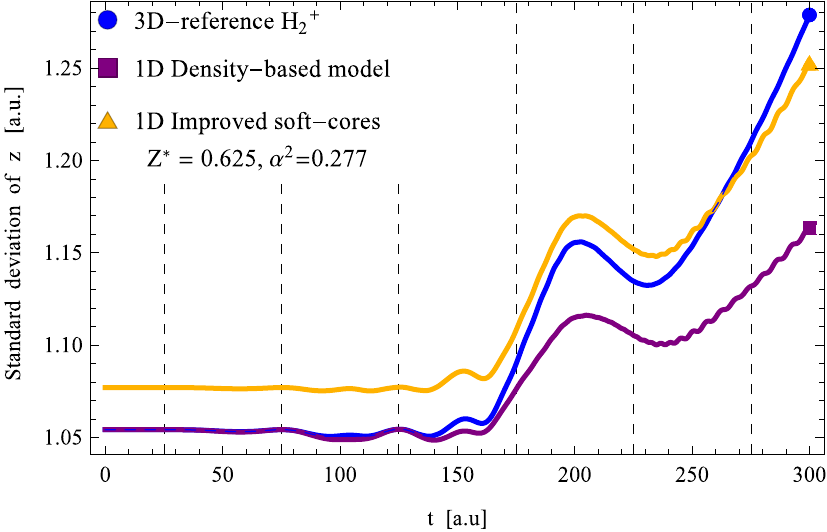}\hspace{0.5cm}\includegraphics[width=1\columnwidth]{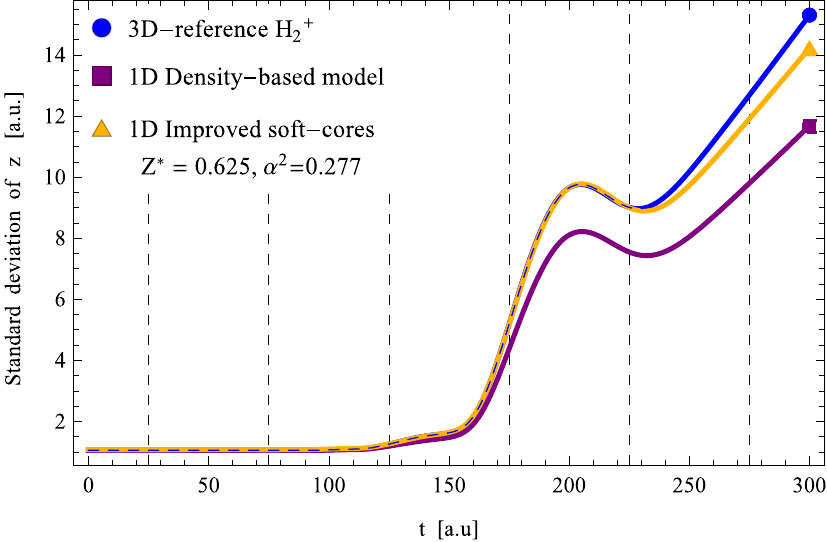}

\begin{raggedright} \hspace{4.5cm}(e)\hspace{8.7cm}(f)

\end{raggedright}

\includegraphics[width=1\columnwidth]{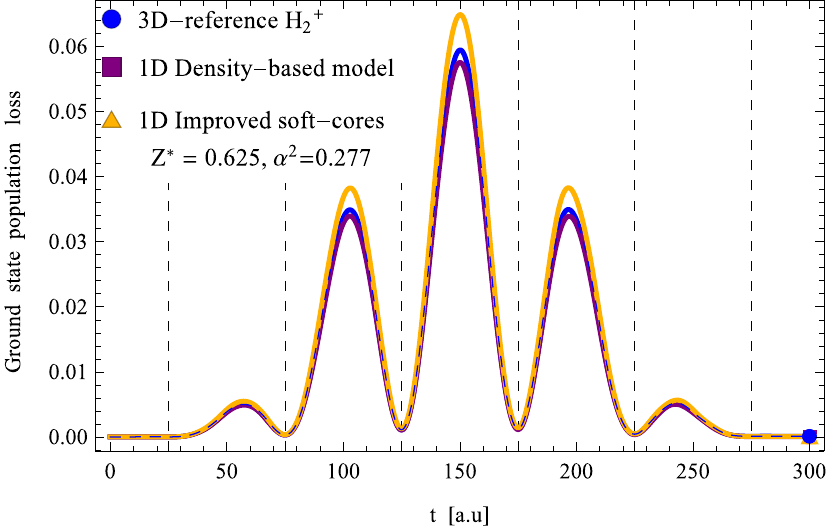}\hspace{0.5cm}\includegraphics[width=1\columnwidth]{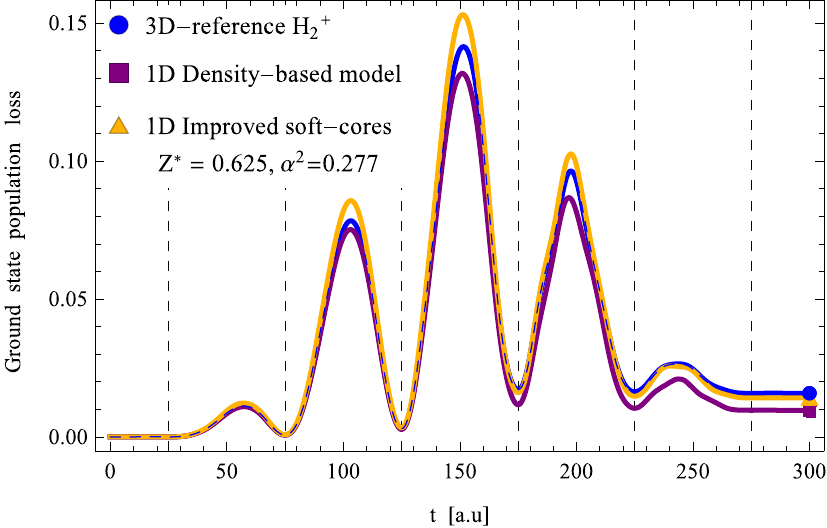}

\protect\protect\protect\caption{Results for a static hydrogen molecular ion with $d=2.0$, driven by the    {near-infrared} laser pulse with $F=0.1$ (left panels) and $F=0.15$ (right panels). We plot the time-dependence
of the mean values $\left\langle z\right\rangle (t)$ in (a)-(b) and
the standard deviations $\sigma_{z}(t)$ in (c)-(d) and the ground state
population losses $g(t)$ in (e)-(f). Results of the corresponding 3D simulations are plotted in blue.}

\label{fig:SinPulse3_D20_H2+} 
\end{figure*}

We present
the results for  $\text{H}_{2}^{+}$  driven by a mid-infrared  pulse of $F=0.15$ 
in the left panels of Fig. \ref{fig:IR_SinPulse1_H2+_He}. 
Comparing these plots  
to those in Fig. \ref{fig:SinPulse3_D20_H2+}, we see that the improved 1D soft-core potential 
gives again a better model of the 3D process than the density-based 1D potential does.
However, the mid-infrared pulse creates  somewhat different dynamics with respect to the near-infrared pulse: 
although $\left\langle z\right\rangle (t)$ in Fig. \ref{fig:SinPulse3_D20_H2+} (a) yet follows the near-infrared pulse shape,  
the $\left\langle z\right\rangle (t)$ in Fig. \ref{fig:IR_SinPulse1_H2+_He} (a) deviates from the mid-infrared pulse shape considerably more than 
expected based on 
Fig. \ref{fig:SinPulse3_D20_H2+} (b), 
due to the longer period.
The longer period also enables  the continuum wavepackets to spread for a longer time,
 thus the  $\sigma_{z}(t)$ has considerably increased values  in Fig. \ref{fig:IR_SinPulse1_H2+_He} (c).
On the other hand, the peaks of the ground state population loss in Fig. \ref{fig:IR_SinPulse1_H2+_He} (e) 
are  in accordance both in magnitude and in timing with those in Fig. \ref{fig:SinPulse3_D20_H2+} (e) and (f), 
but the final value of  of $g(t)$ in Fig. \ref{fig:IR_SinPulse1_H2+_He} (e) is considerably less than that in Fig. \ref{fig:SinPulse3_D20_H2+} (f).
This latter shows that strong-field ionization from the double-well potential of $\text{H}_{2}^{+}$, which has interesting internal and strong-field dynamics \cite{CarlaMorissonFaria_NJP_2019}, 
is very sensitive to the peak value of the laser electric field which is  ca. 20\% lower for the mid-infrared than for the near-infrared pulse.

\begin{figure*}
\begin{raggedright} \hspace{4.5cm}(a)\hspace{8.7cm}(b)

\end{raggedright}

\includegraphics[width=1\columnwidth]{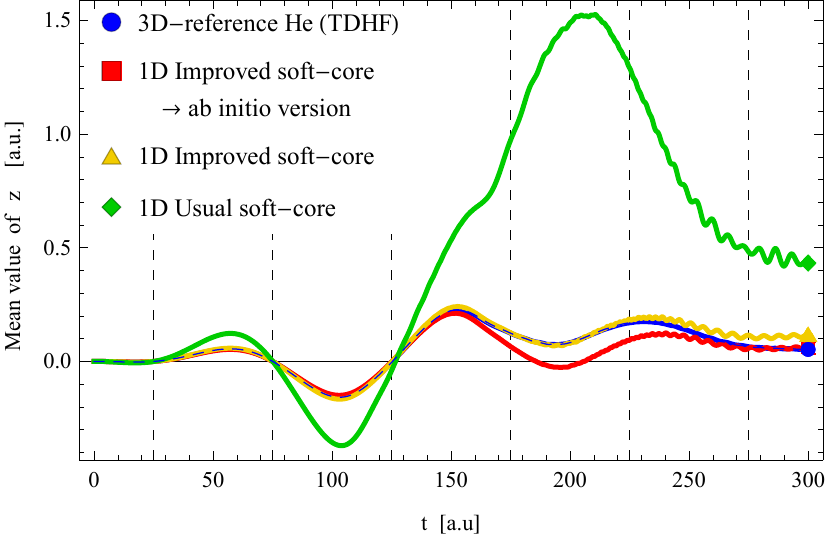}\hspace{0.5cm}\includegraphics[width=1\columnwidth]{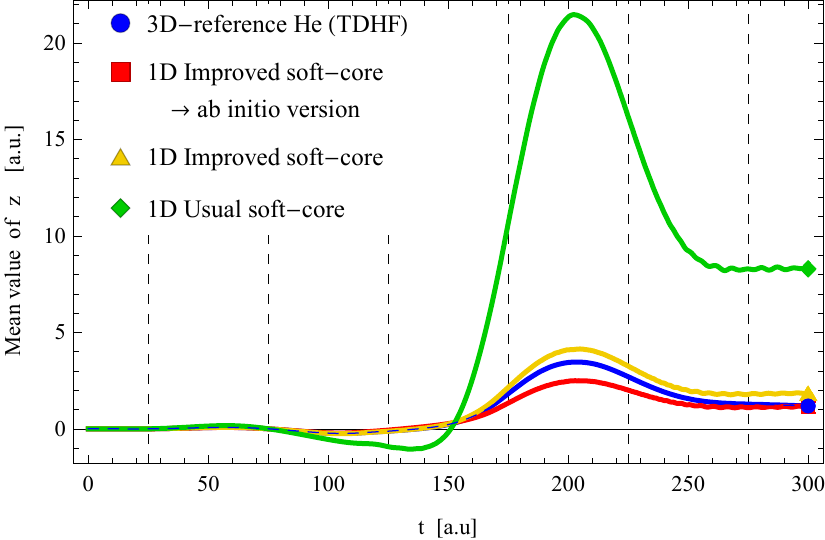}

\begin{raggedright} \hspace{4.5cm}(c)\hspace{8.7cm}(d)

\end{raggedright}

\includegraphics[width=1\columnwidth]{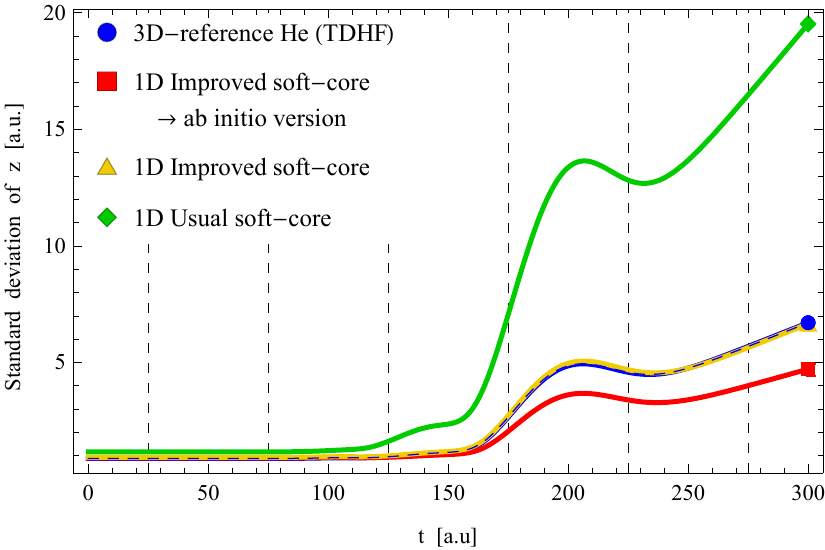}\hspace{0.5cm}\includegraphics[width=1\columnwidth]{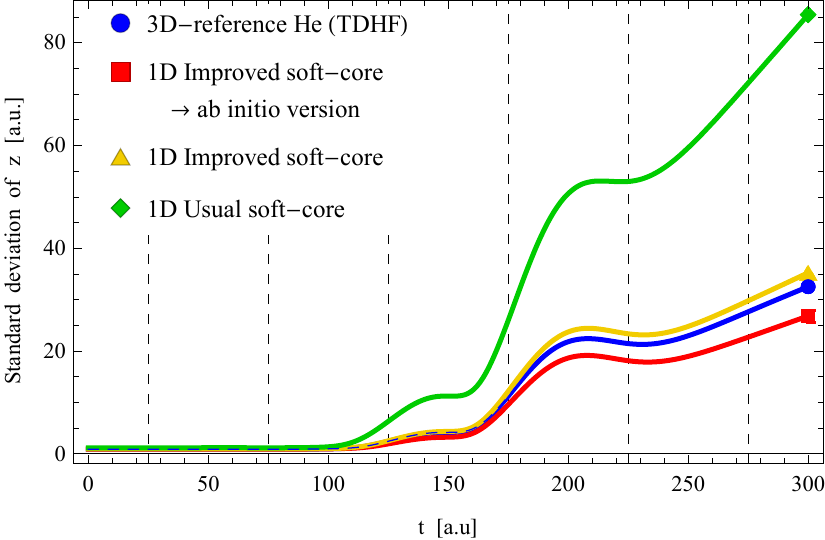}

\begin{raggedright} \hspace{4.5cm}(e)\hspace{8.7cm}(f)

\end{raggedright}

\includegraphics[width=1\columnwidth]{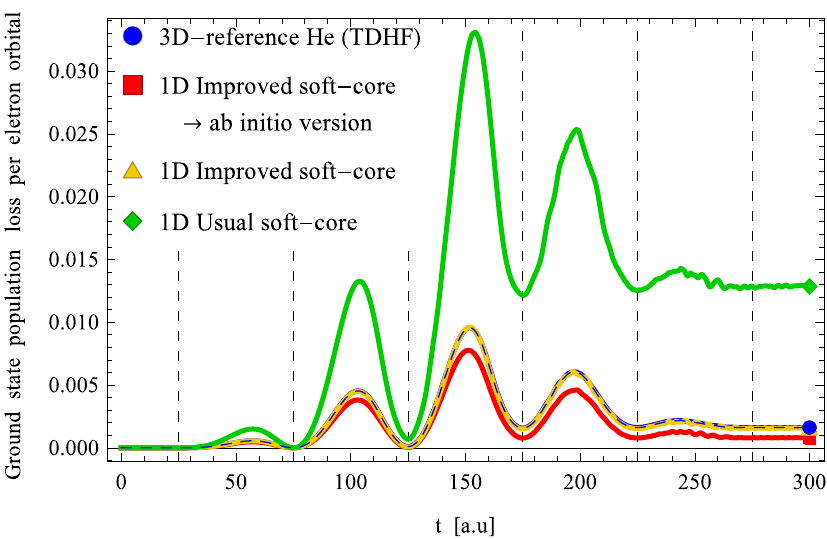}\hspace{0.5cm}\includegraphics[width=1\columnwidth]{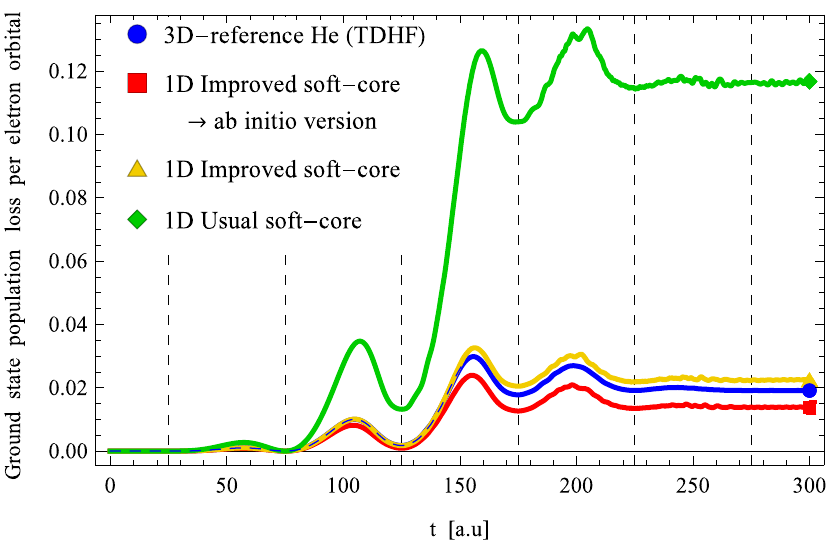}

\protect\protect\protect\caption{The time-dependence of the mean values
$\left\langle z\right\rangle (t)$ in (a)-(b) and the standard deviations
$\sigma_{z}(t)$ in (c)-(d) and the ground state population losses $g(t)$
in (e)-(f) for a helium atom driven by the    {near-infrared} laser pulse with $F=0.15$ (left
panels) and $F=0.20$ (right panels). Results of the corresponding
3D simulations are plotted in blue.}

\label{fig:SinPulse3_He} 
\end{figure*}

\begin{figure*}
\begin{raggedright} \hspace{4.5cm}(a)\hspace{8.7cm}(b)

\end{raggedright}

\includegraphics[width=1\columnwidth]{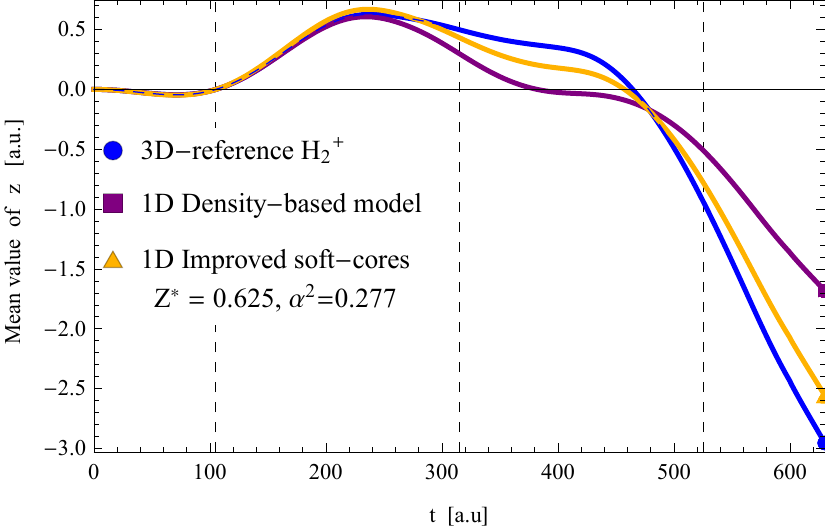}\hspace{0.5cm}\includegraphics[width=1\columnwidth]{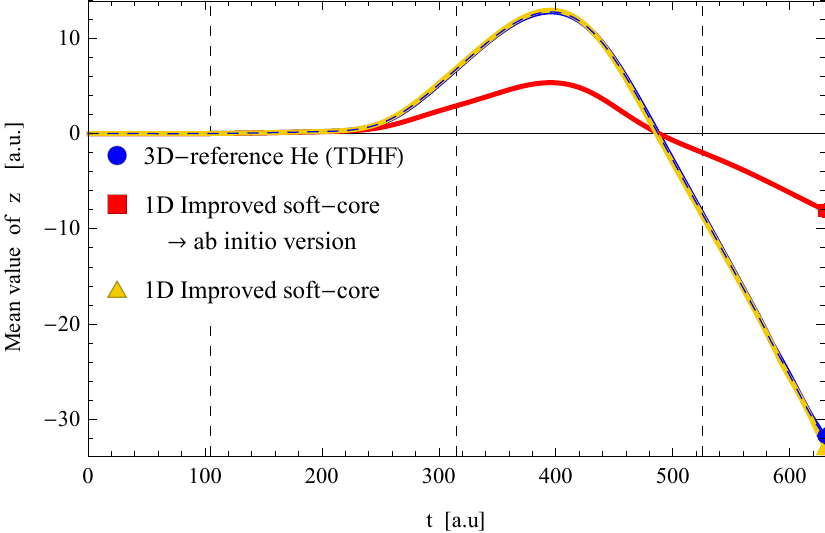}

\begin{raggedright} \hspace{4.5cm}(c)\hspace{8.7cm}(d)

\end{raggedright}

\includegraphics[width=1\columnwidth]{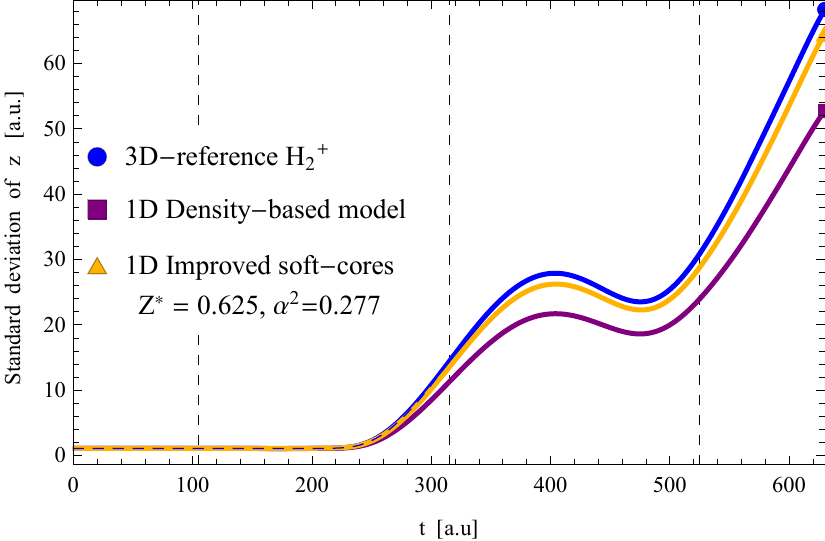}\hspace{0.5cm}\includegraphics[width=1\columnwidth]{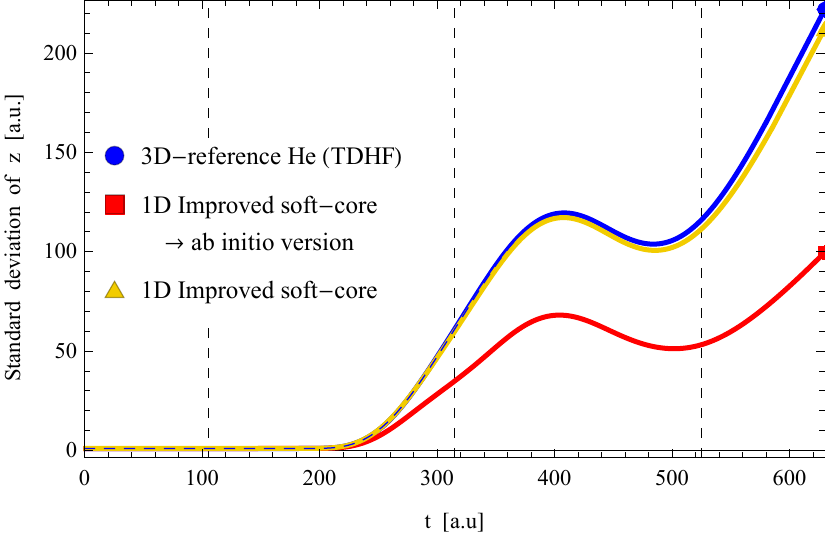}

\begin{raggedright} \hspace{4.5cm}(e)\hspace{8.7cm}(f)

\end{raggedright}

\includegraphics[width=1\columnwidth]{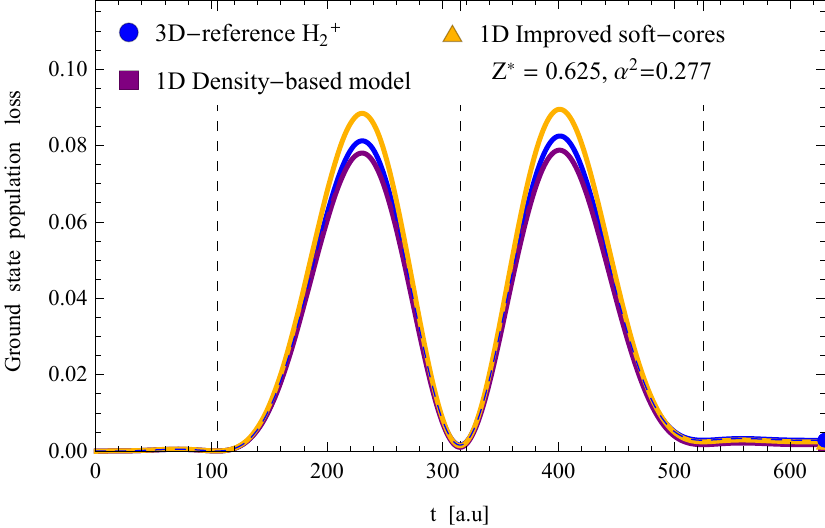}\hspace{0.5cm}\includegraphics[width=1\columnwidth]{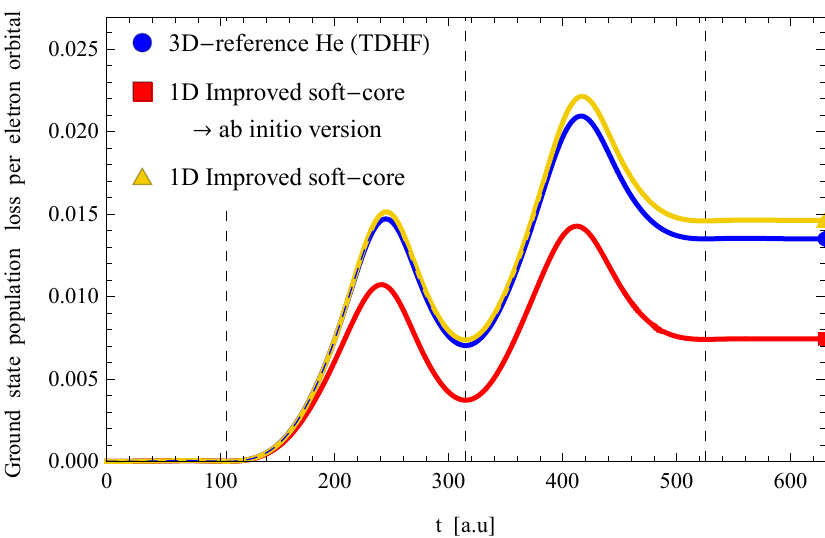}

\protect\protect\protect\caption{
   {Results for a static hydrogen molecular ion with $d=2.0$
(left panels), and for a helium atom (right panels), driven
by the mid-infrared laser pulse with
 $F=0.15$ for $\text{H}_{2}^{+}$ and $F=0.2$ for He. 
We plot the time-dependence of the mean value $\left\langle z\right\rangle (t)$
in (a)-(b), the standard deviation $\sigma_{z}(t)$ in (c)-(d)
and the ground state population loss $g(t)$ in (e)-(f). The color coding of the curves in
the left and right panels corresponds to those in Fig. \ref{fig:SinPulse3_D20_H2+}
and Fig. \ref{fig:SinPulse3_He}, respectively.}}

\label{fig:IR_SinPulse1_H2+_He} 
\end{figure*}

We present
the results for  He driven by a mid-infrared  pulse with $F=0.2$ 
in the right panels of Fig. \ref{fig:IR_SinPulse1_H2+_He}, which are to be compared to  the right panels of
Fig. \ref{fig:SinPulse3_He}. 
The  1D results corresponding to $V_{0,\mathrm{Sc}}(z)$ with  $\alpha^{2}=0.0694$ give even better results than those for the near-infrared pulse, 
but the "ab initio" parameter for $V_{0,\mathrm{Sc}}(z)$ is clearly less accurate for the mid-infrared pulse. This suggests that the exact  match of the 
1D ground state energy with the 3D ground state energy becomes  more important with increasing laser wavelength.
The difference in the shapes of the $\left\langle z\right\rangle (t)$ for the near- and the mid-infrared pulse is clearly due to the different 
 pulse shapes
  and the increased wavelength.
The ground state population loss is very similar to Fig. \ref{fig:SinPulse3_He} (f), both in behaviour and in magnitude,
which is readily explained by taking into account that the increasing effect of the longer period of the mid-infrared pulse
is  largely compensated by the actually ca. 20\% less peak electric field strength.  
However, the  longer period of the mid-infrared pulse does considerably
 increase the values of   the  $\left\langle z\right\rangle (t)$ and
$\sigma_{z}(t)$  curves, since the continuum part of the wave function has  more time to travel and spread.

Summarizing this section, the carefully parametrized 1D soft-core model potentials work very well also in the case of a mid-infrared laser pulse. 
Note also, that the small oscillations visible on some of the 1D curves in Figs.  \ref{fig:SinPulse3_D20_H2+}-\ref{fig:SinPulse3_He_F0.25}  are absent now,
 due to the mid-infrared pulse shape which
causes  only one rescattering.

\section{Discussion and conclusions}

In this paper we have shown that the density-based model potentials,
developed in \cite{majorosi2018densitymodel} for strong field simulations
in single-active-electron atoms can be extended to two-electron systems
and simple molecules. Our results show that the modeling based on
low dimensional Coulomb asymptotes selected by the reduced density
works also for two-electron problems, not just for single-active electron
systems. The molecular potentials we built in this way fulfilled the
requirement that the 1D model should recover a quantitatively correct
ionization response compared to the respective 3D molecular system
under the influence of a linearly polarized external laser field.
This improvement is mainly due to the accurate match of the 1D and 3D ground state energy and 
the correct 1D asymptotics which provides better 1D continuum states.

For H$_{2}^{+}$ one of the molecular potentials has been obtained
by using the reduced density according to Eq. \eqref{eq:1d_density_pot_mol}.
The other choice for this system was a sum of two soft-core Coulomb
potentials as given by Eq. \eqref{eq:1d_pot_soft} with appropriate
values of the fitting parameters $Z^{\ast}$ and $\alpha^{2}$. For
the two-electron systems He and H$_{2}$ the potentials were built
by combining 1D improved soft-core Coulomb potentials from the
corresponding single-electron density-based models. Here we have used
two different sets of parameters to simultaneously reproduce the correct
density based Coulomb asymptotics and ground state energies. 
We compared the results of numerical strong-field simulations for
the complete 3D systems with our 1D molecular models excited with
the same linearly polarized laser field. 
We have shown that our simpler
1D constructions provide impressive accuracy, for being 1D models, for
the helium atom and the hydrogen molecular ion, driven by the near-infrared and even by the mid-infrared pulse. 
These two models
performed exceptionally well, especially if the strong-field ionization response
was relatively weak. For the case of the hydrogen molecule the correct asymptotic
values of the potential turned out to be near to that of the hydrogen
molecular ion with the same intermolecular distance. 
Experimental
interest in strong field and attosecond processes of He \cite{shafir2012tunneling, chen2012heliumabsorb, stooss2018strongfieldresponse} inspired
us to calculate high-order-harmonic generation spectra \ by using
our 1D model for He. 
It turned out that quantitatively correct spectra
could be recovered with a simple scaling for different external electric
fields.

These results overall provide more possibilities to explore. One can extend
the description of the molecular systems and model them under strong-field
circumstances with moving nuclei. In order to include the motion of
the nuclei into the simulation using e.g. the Born-Oppenheimer or
the Ehrenfest approximations \cite{BOOK_TDDFT_ULLRICH,BOOK_ATOMS_MOLECULES_BRANSDEN_2003},
the determination of $Z^{\ast}(d)$ and $\alpha^{2}(d)$ is required
beforehand numerically. Such a $d$ dependent molecular model potential
seems to be a promising way of modeling the true molecular dynamics
in strong-field scenarios with electron wave functions in one spatial
variable, which would make it especially effective. It is also a possibility
to model a linear chain of atomic cores in 1D by reducing the proper
3D density in advance. Regarding the multiple electron systems,
one can apply the time-dependent Hartree-Fock, or multiconfigurational
Hartree-Fock approach \cite{beck2000mctdhbig} to the 1D helium and
hydrogen systems directly, which may enable massive performance gain,
while providing quantitatively correct reduced dynamics and high-order-harmonic
spectra. This would also enable to efficiently perform low dimensional
strong-field calculations in gas-jets, or statistical mixtures.

\begin{acknowledgments}
The authors thank Péter Földi, Katalin Varjú and Sándor Varró  for stimulating discussions.
This research was performed in the framework of the project Nr. GINOP-2.3.2-15-2016-00036 titled Development and application of multimodal optical nanoscopy methods in life and material sciences.
The project has been also supported by the European Union, co-financed by the European Social Fund, Grant
No. EFOP-3.6.2-16-2017-00005.
Partial support by the ELI-ALPS project is also acknowledged. 
The ELI-ALPS project (GINOP-2.3.6-15-2015-00001) is supported 
by the European Union and co-financed by the European Regional Development Fund.
\end{acknowledgments}

\bibliographystyle{unsrtnat}
\bibliography{0Bibliography,2Bibliography,3Bibliography}

\end{document}